\documentclass[11pt]{article}
\usepackage{times}
\usepackage{mathfont}
\usepackage{graphicx}
\usepackage{cite}
\usepackage{url}
\def\max{\mathop{\rm max}}

\mathcode`O="724F

\DeclareSymbolFont{lasy}{U}{lasy}{m}{n}
\let\Box\undefined
\DeclareMathSymbol\Box{0}{lasy}{"32}

\newtheorem{lemma}{Lemma}
\newtheorem{corollary}{Corollary}
\newtheorem{theorem}{Theorem}

\newcommand{\qed}{~$\Box$\medbreak}
\newenvironment{proof}{\noindent{\bf Proof: }}{\qed}

\begin{document}

\title{3-Coloring in Time $O(1.3289^{\textstyle n})$}

\author{Richard Beigel\thanks{
Dept. Elect. Eng. \& Comp. Sci. (m/c 154),
Univ. of Illinois, Chicago, 851 S. Morgan St., Fl. 11,
Chicago, IL 60607-7053.
Email: {\tt beigel@uic.edu}.
Supported in part by
NSF grants CCR-8958528 and CCR-9415410.}\\
{\small Univ. of Illinois, Chicago}
\and
David Eppstein\thanks{Dept. Inf. \& Comp. Sci., Univ. of
California, Irvine, CA 92697-3425.  Email: {\tt eppstein@ics.uci.edu}.
Supported in part by NSF grant
CCR-9258355 and by matching funds from Xerox Corp.} \\
{\small Univ. of California, Irvine}}

\date{ }
\maketitle

\begin{abstract}
We consider worst case time bounds for NP-complete problems including
3-SAT, 3-coloring, 3-edge-coloring, and 3-list-coloring.  Our
algorithms are based on a constraint satisfaction (CSP) formulation of
these problems.  3-SAT is equivalent to $(2,3)$-CSP
while the other problems above are special cases of $(3,2)$-CSP; there is
also a natural duality transformation from $(a,b)$-CSP to $(b,a)$-CSP.  We
give a fast algorithm for $(3,2)$-CSP and use it to improve the
time bounds for solving the other problems listed above.
Our techniques involve a mixture of Davis-Putnam-style backtracking with
more sophisticated matching and network flow based ideas.
\end{abstract} 

\section{Introduction}

There are many known NP-complete problems including such important graph
theoretic problems as coloring and independent sets.
Unless P=NP, we know that no polynomial time algorithm for these
problems can exist, but that does not obviate the need to solve them as
efficiently as possible, indeed the fact that these problems are hard
makes efficient algorithms for them especially important.

We are interested in this paper in worst case analysis of algorithms
for 3-coloring, a basic NP-complete problem.
We will also discuss other related problems including
3-SAT, 3-edge-coloring and 3-list-coloring.

Our algorithms for these problems are based on the following simple
idea: to find a solution to a 3-coloring problem, it is not necessary to
choose a color for each vertex (giving something like $O(3^n)$ time).
Instead, it suffices to only partially solve the problem by restricting
each vertex to two of the three colors.  We can then test whether the
partial solution can be extended to a complete coloring in polynomial
time (e.g. as a 2-SAT instance).  This idea applied naively already
gives a simple $O(1.5^n)$ time randomized algorithm; we improve this by
taking advantage of local structure (if we choose a color for one
vertex, this restricts the colors of several neighbors at once).  It
seems likely that our idea of only searching for a partial solution
can be applied to many other combinatorial search problems.

If we perform local reductions as above in a 3-coloring problem,
we eventually reach a situation in which some uncolored vertices
are surrounded by partially colored neighbors, and we run out of good
local configurations to use.  To avoid this problem, we translate our
3-coloring problem to one that also generalizes the other problems
listed above: {\em constraint satisfaction} (CSP).
In an $(a,b)$-CSP instance, we are given a collection of $n$ variables,
each of which can be given one of $a$ different colors.
However certain color combinations are disallowed: we also have input
a collection of $m$ {\em constraints}, each of which forbids one coloring
of some $b$-tuple of variables.
Thus 3-satisfiability is exactly $(2,3)$-CSP,
and 3-coloring is a special case of $(3,2)$-CSP in which
the constraints disallow adjacent vertices from having the same color.

As we show, $(a,b)$-CSP instances can be transformed in certain
interesting and useful ways: in particular, one can transform
$(a,b)$-CSP into $(b,a)$-CSP and vice versa, one can transform
$(a,b)$-CSP into $(\max(a,b),2)$-CSP, and in any $(a,2)$-CSP instance one
can eliminate variables for which only two colors are allowed, reducing
the problem to a smaller one of the same form.
Because of this ability to eliminate partially colored variables
immediately rather than saving them for a later 2-SAT instance, we can
solve a $(3,2)$-CSP instance without running out of good local
configurations.

Our actual algorithm solves $(3,2)$-CSP by applying such reductions only
until we reach instances with a certain simplified structure, which can
then be solved in polynomial time as an instance of graph matching.
We further improve our time bound for graph 3-vertex-coloring by using
methods involving network flow to find a large set of good local
reductions which we apply before treating the remaining problem as a
$(3,2)$-CSP instance. And similarly, we solve 3-edge-coloring by using
graph matching methods to find a large set of good
local reductions which we apply before treating the remaining problem as a
3-vertex-coloring instance.

\subsection{New Results}

We show the following:

\begin{itemize}
\item A $(3,2)$-CSP instance with $n$ variables can be solved in worst
case time $O(1.3645^n)$, independent of the number of constraints.
We also give a very simple randomized algorithm for solving this
problem in expected time $O(n^{O(1)} 2^{n/2})\approx O(1.4142^n)$.

\item A $(d,2)$-CSP instance with $n$ variables and $d>3$ can be solved by
a randomized algorithm in expected time $O((0.4518d)^n)$.

\item 3-coloring in a graph of $n$ vertices can be solved in time
$O(1.3289^n)$, independent of the number of edges in the graph.

\item 3-list-coloring (graph coloring given a list at each vertex of
three possible colors chosen from some larger set) can be solved in time
$O(1.3645^n)$, independent of the number of edges.

\item 3-edge-coloring in an $n$-vertex graph can be solved in time
$O(2^{n/2})$, again independent of the number of edges.

\item 3-satisfiability of a formula with $t$ 3-clauses can be solved in
time $O(n^{O(1)}+1.3645^t)$, independent of the number of variables or
2-clauses in the formula.
\end{itemize}

Except where otherwise specified, $n$ denotes the number of
vertices in a graph or variables in a SAT or CSP instance, while
$m$ denotes the number of edges in a graph, constraints in an CSP
instance, or clauses in a SAT problem.

\subsection{Related Work}

There is a growing body of papers on worst case analysis of algorithms
for NP-hard problems. Several authors have described
algorithms for maximum independent
sets~\cite{Bei-SODA-99,CheKanJia-WG-99,Jia-TC-86,ParRapRes-HPA-99,
Rob-Algs-86,ShiTom-SCJ-90,TarTri-SJC-77};
the best of these is Robson's~\cite{Rob-Algs-86}, which takes time
$O(1.2108^n)$.  Others have described algorithms for Boolean
formula
satisfiability~\cite{Dan-JSM-83,DanHir-TR-00,DavPut-JACM-60,
GraHirNie-SAT-00,Hir-SODA-98,Luc-FC-84,KulLuc-ms-94,Kul-TCS-99,
MonSpe-DAM-85,PatPudSak-FOCS-98,Rod-AISMC-96,Sch-CSL-92,Sch-FOCS-99};
the best of these satisfiability algorithms are Sch\"oning's, which
solves 3-SAT in expected time $O((4/3)^n)$ \cite{Sch-FOCS-99}, and
Hirsch's, which solves SAT in time $O(1.239^m)$ \cite{Hir-SODA-98}.

For three-coloring, we know of several relevant references.
Lawler~\cite{Law-IPL-76} is primarily concerned with the general chromatic
number, but he also gives the following very simple algorithm for
3-coloring: for each maximal independent set, test whether the
complement is bipartite.  The maximal independent sets can be listed
with polynomial delay~\cite{JohYanPap-IPL-88}, and there are at most
$3^{n/3}$ such sets~\cite{MooMos-IJM-65}, so this algorithm takes time
$O(1.4422^n)$. Schiermeyer~\cite{Sch-WG-93} gives a complicated algorithm
for solving 3-colorability in time $O(1.415^n)$, based on the following
idea: if there is one vertex $v$ of degree $n-1$ then the graph is
3-colorable iff $G-v$ is bipartite, and the problem is easily solved.
Otherwise, Schiermeyer performs certain reductions involving maximal
independent sets that attempt to increase the degree of $G$
while partitioning the problem into subproblems, at least one of which
will remain solvable.  Our $O(1.3289^n)$ bound significantly improves
both of these results.

There has also been some related work on approximate or heuristic
3-coloring algorithms. Blum and Karger~\cite{BluKar-IPL-97} show that
any 3-chromatic graph can be colored with $\tilde{O}(n^{3/14})$ colors in
polynomial time. Alon and Kahale~\cite{AloKah-SJC-97} describe a technique
for coloring random 3-chromatic graphs in expected polynomial time, and
Petford and Welsh~\cite{PetWel-DM-89} present a randomized algorithm for
3-coloring graphs which also works well empirically on random graphs
although they prove no bounds on its running time. Finally,
Vlasie~\cite{Vla-TAI-95} has described a class of instances which are
(unlike random 3-chromatic graphs) difficult to color.

Very recently, Sch\"oning~\cite{Sch-FOCS-99} has described a simple and
powerful randomized algorithm for $k$-SAT and more general constraint
satisfaction problems, including the CSP instances that we use in our
solution of 3-coloring.  However, for $(d,2)$-CSP, Sch\"oning notes that
his method is not as good as a randomized approach based on an idea from
our previous conference paper~\cite{BeiEpp-FOCS-95}: simply
choose a random pair of values for each variable and solve the resulting
2-SAT instance in polynomial time.  The table below compares the
resulting $(d/2)^n$ bound with our new results; an entry with value $x$
in column $d$ indicates a time bound of
$O(x^n)$ for $(d,2)$-CSP.

\begin{table}[h]
\begin{center}
\begin{tabular}{|l|l|l|l|l|} \hline
&$d=3$&$d=4$&$d=5$&$d=6$\\ \hline
Sch\"oning~\cite{Sch-FOCS-99}&1.5&2&2.5&3\\ \hline
New results&1.3645&1.8072&2.2590&2.7108\\ \hline
\end{tabular}
\end{center}
\end{table}

We were unable to locate prior work on worst case edge coloring. Since
any 3-edge-chromatic graph has at most $3n/2$ edges, one can transform
the problem to 3-vertex-coloring at the expense of increasing $n$ by a
factor of $3/2$.  If we applied our vertex coloring algorithm we would
then get time $O(1.5319^n)$ which is significantly improved by the bound
stated above.

It is interesting that, historically, until the work of
Sch\"oning~\cite{Sch-FOCS-99}, the time bounds for 3-coloring have been
smaller than those for 3-satisfiability (in terms of the number of
vertices or variables respectively).  Sch\"oning's $O((4/3)^n)$ bound for
3-SAT reversed this pattern by being smaller than the previous
$O(1.3443^n)$ bound for 3-coloring from our
1995 conference paper~\cite{BeiEpp-FOCS-95}.  The present work restores
3-coloring to a smaller time bound than 3-SAT.

\section{Constraint Satisfaction Problems}

We now describe a common generalization of satisfiability and graph
coloring as a {\em constraint satisfaction problem}
(CSP)~\cite{Kum-AIM-92,Sch-FOCS-99}. We are given a collection of $n$
variables, each of which has a list of possible colors allowed.
We are also given a collection of $m$ {\em constraints},
consisting of a tuple of variables and a color for each variable.
A constraint is {\em satisfied} by a coloring if not every variable in
the tuple is colored in the way specified by the constraint. We would
like to choose one color from the allowed list of each variable, in a way
not conflicting with any constraints.

\begin{figure}[t]
$$\includegraphics[width=2.6in]{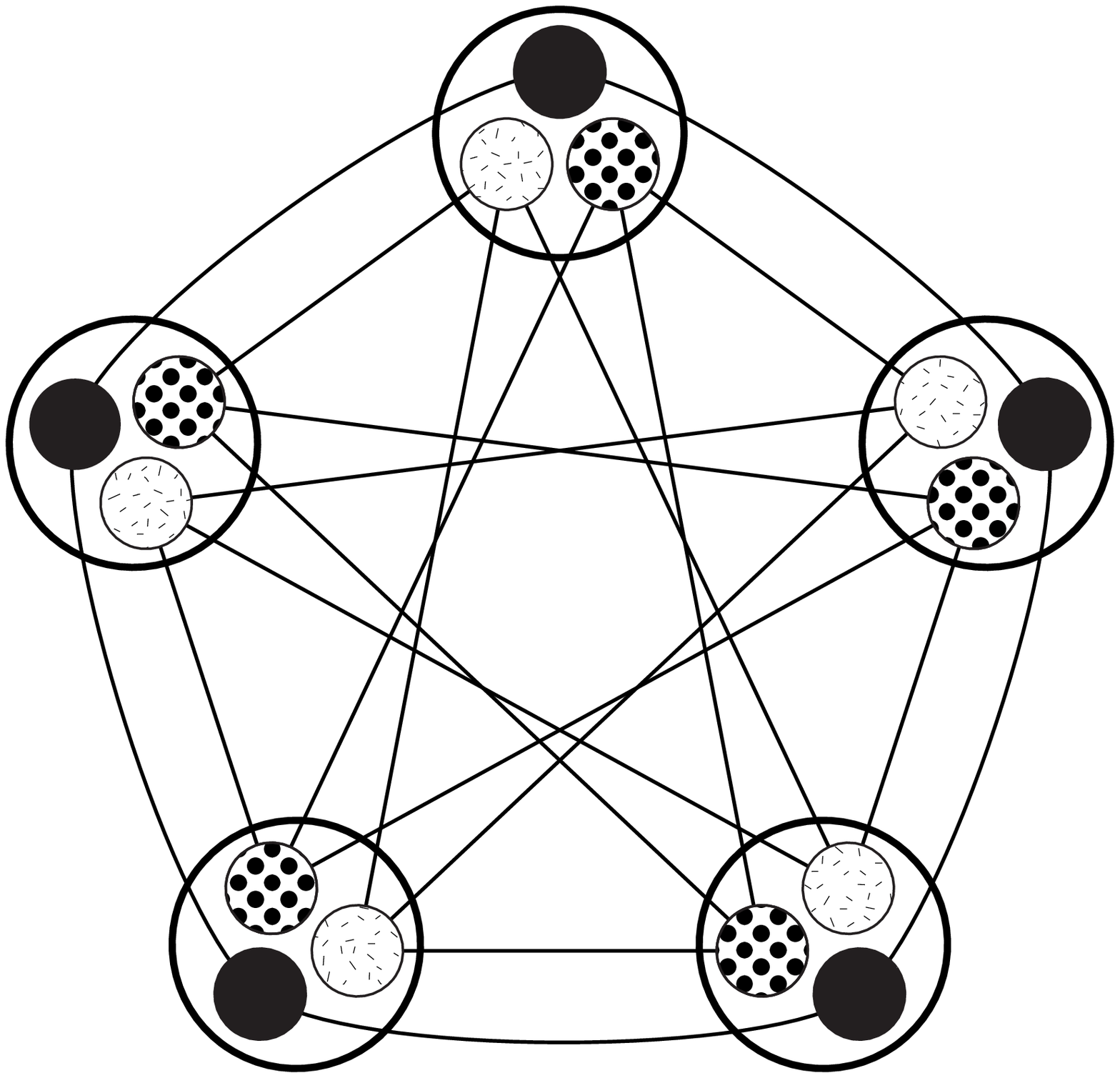}\qquad
\includegraphics[width=2.6in]{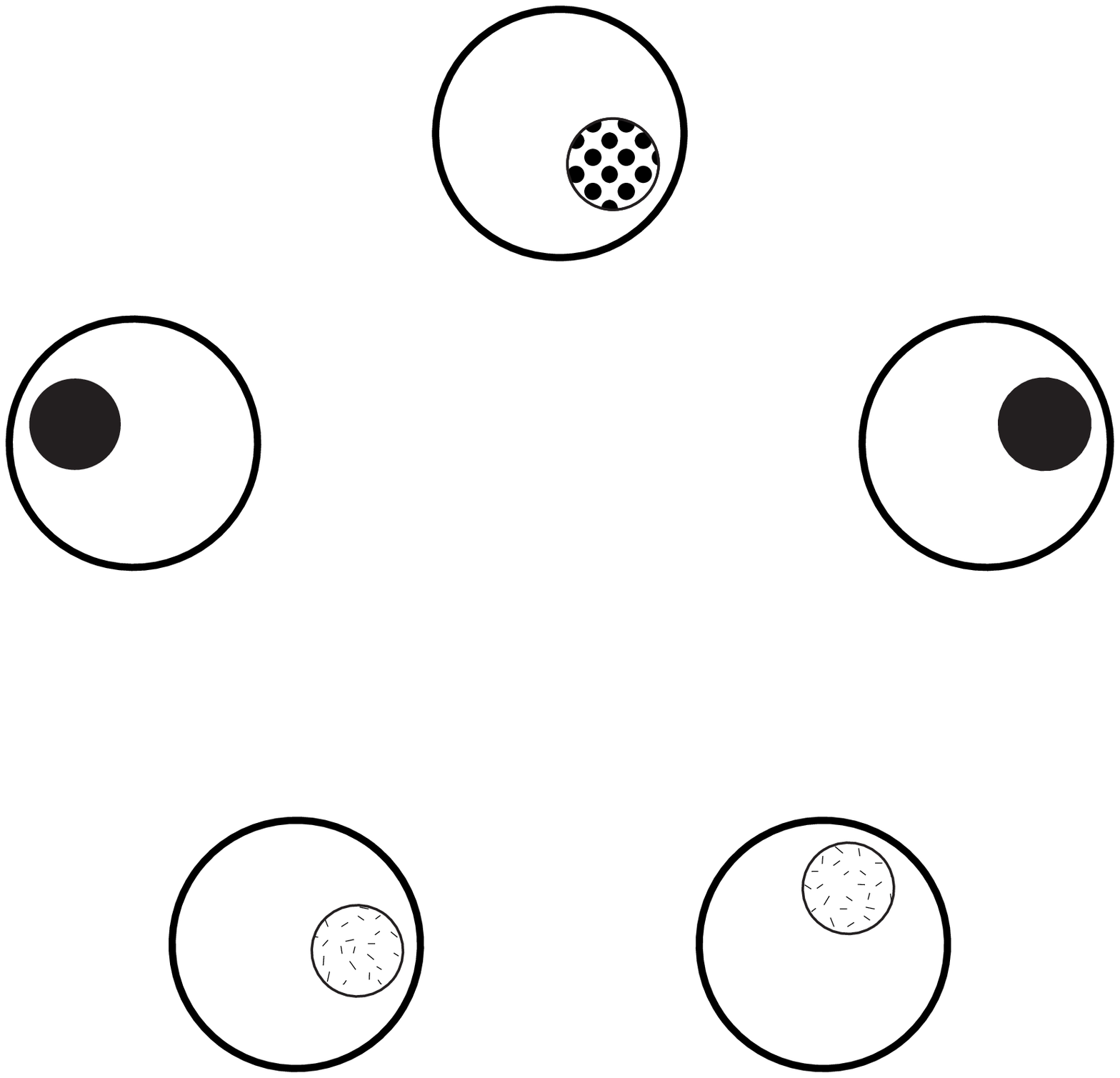}$$
\caption{Example $(3,2)$-CSP instance with five variables and twenty
constraints (left), and a solution of the instance (right).}
\label{fig:32sss}
\end{figure}

\begin{figure}[t]
$$\includegraphics[width=2.6in]{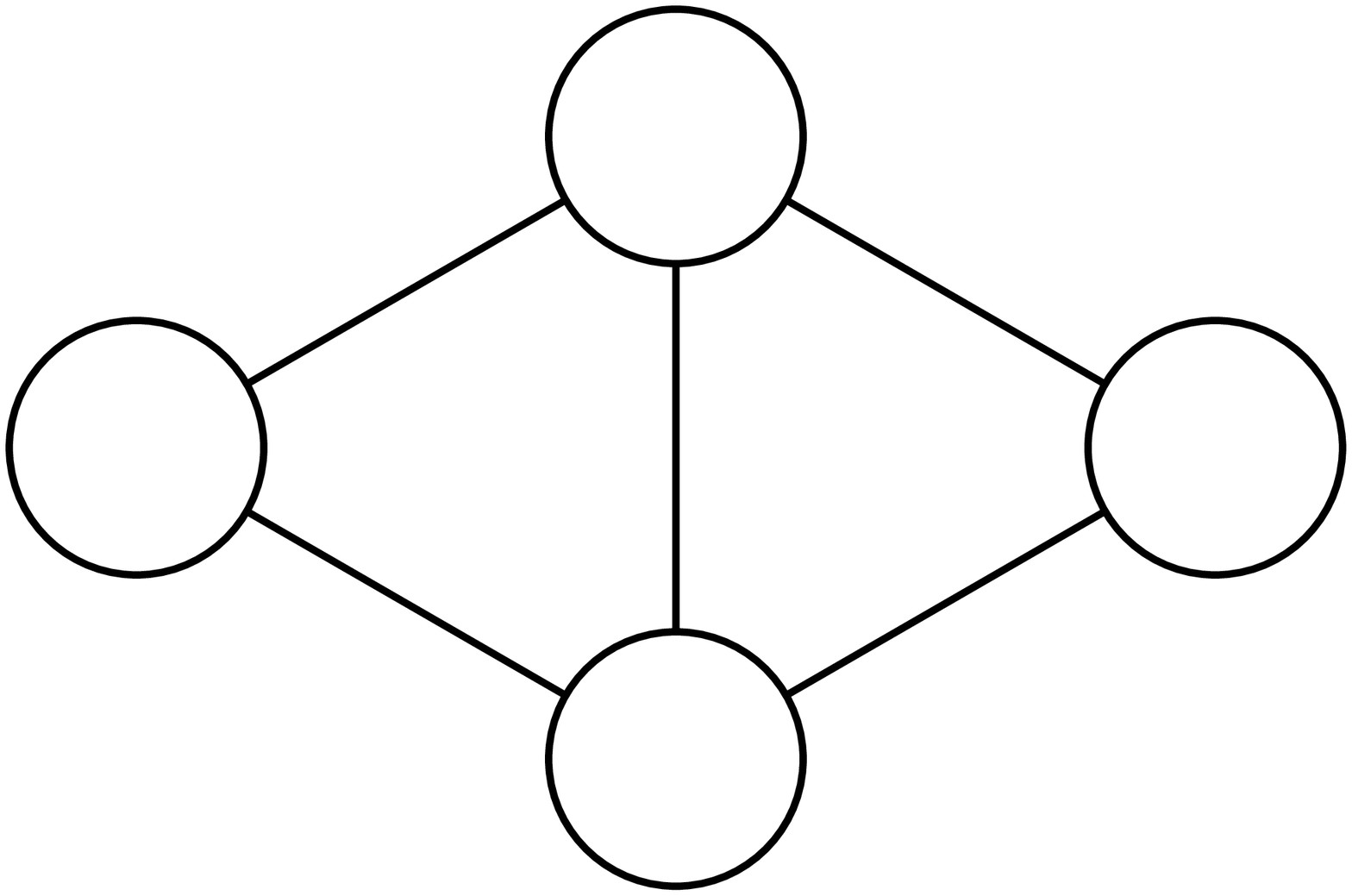}\qquad
\includegraphics[width=2.6in]{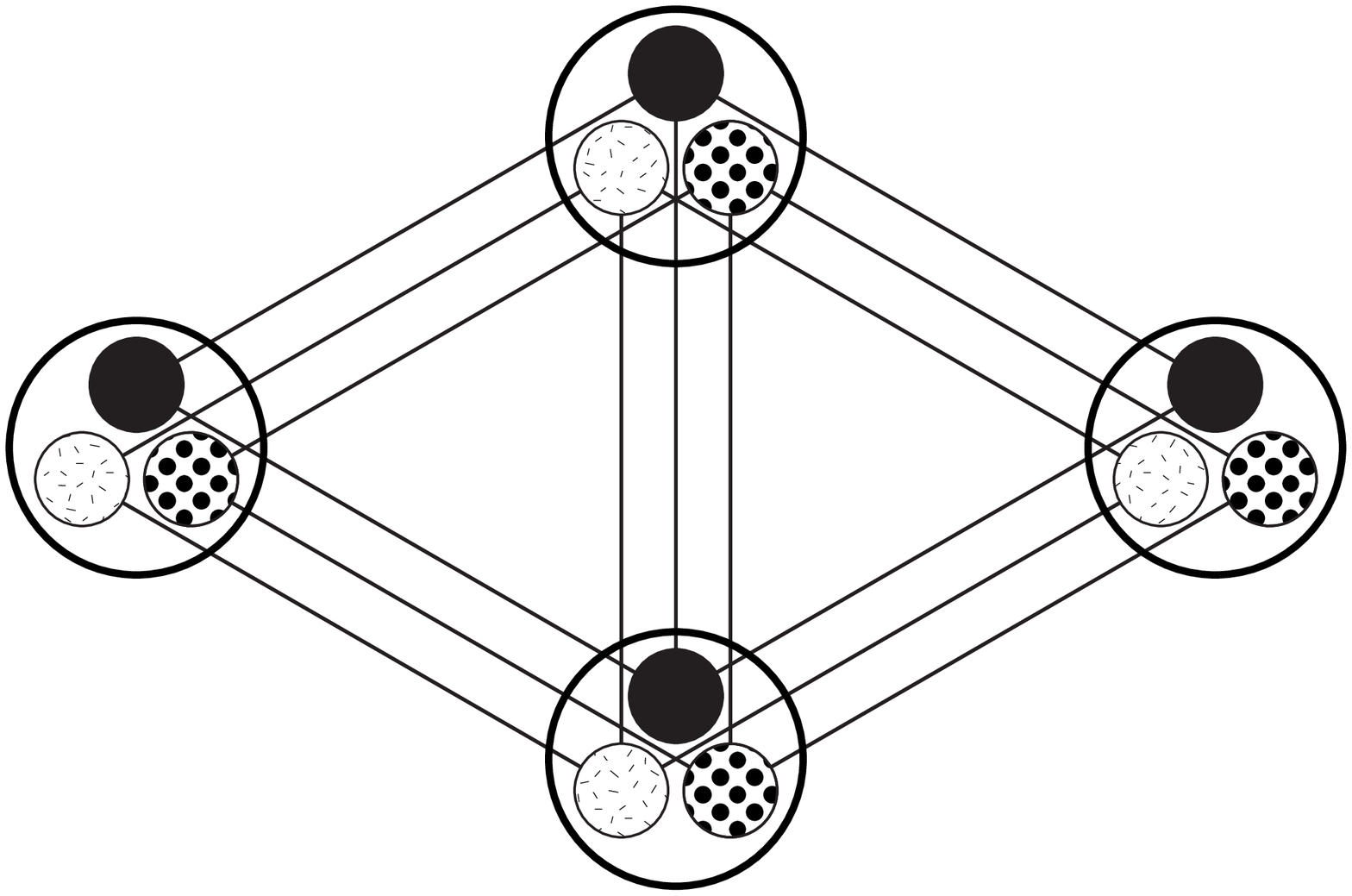}$$
\caption{Example 3-coloring instance (left)
and translation into a $(3,2)$-CSP instance (right).}
\label{fig:gc2sss}
\end{figure}

For instance, 3-satisfiability can easily be expressed in this form.
Each variable of the satisfiability problem may be colored (assigned the
value) either {\em true} ($T$) or {\em false}
($F$). For each clause like $(x_1 \vee x_2 \vee \neg{x}_3)$,
we make a constraint $((v_1,F),(v_2,F),(v_3,T))$.
Such a constraint is satisfied if and only if at least one of the
corresponding clause's terms is true.

In the {\em $(a,b)$-CSP} problem,
we restrict our attention to instances in which each variable has at most
$a$ possible colors and each constraint involves at most $b$ variables.
The CSP instance constructed above from a 3-SAT instance is then
a $(2,3)$-CSP instance, and in fact 3-SAT is easily seen to be
equivalent to $(2,3)$-CSP.

In this paper, we will concentrate our
attention instead on $(3,2)$-CSP and $(4,2)$-CSP.  We can represent a
$(d,2)$-CSP instance graphically, by interpreting each variable as a
vertex containing up to $d$ possible colors, and by drawing edges
connecting incompatible pairs of vertex colors (Figure~\ref{fig:32sss}). 
Note that this graphical structure is not actually a graph, as the edges
connect colors within a vertex rather than the vertices themselves.
However, graph $3$-colorability and graph $3$-list-colorability
can be translated directly to a form of $(3,2)$-CSP: we keep the
original vertices of the graph and their possible colors, and add up to
three constraints for each edge of the graph to enforce the condition
that the edge's endpoints have different colors
(Figure~\ref{fig:gc2sss}).

Of course, since these problems are all NP-complete, the theory of
NP-completeness provides translations from one problem to the other,
but the translations above are size-preserving and very simple. We will
later describe more complicated translations from 3-coloring and
3-edge-coloring to $(3,2)$-CSP in which the input graph is partially
colored before treating the remaining graph as an CSP instance, leading
to improved time bounds over our pure CSP algorithm.

As we now show, $(a,b)$-CSP instances can be transformed in certain
interesting and useful ways.  We first describe a form of duality
that transforms $(a,b)$-CSP instances into $(b,a)$-CSP instances,
exchanging constraints for variables and vice versa.

\begin{figure}[t]
$$\includegraphics[width=3.5in]{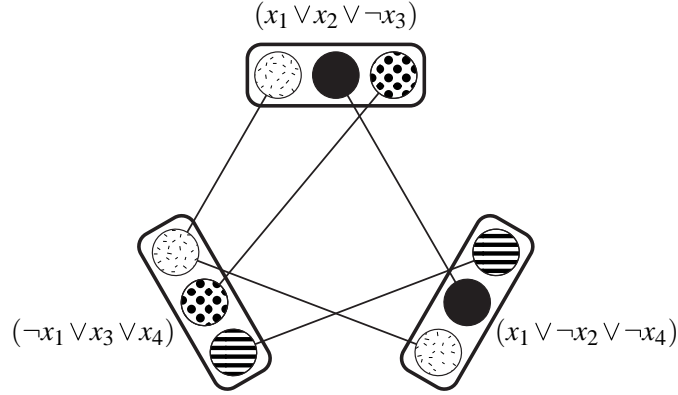}$$
\caption{Translation from 3-SAT (or $(2,3)$-CSP) to $(3,2)$-CSP.}
\label{fig:3s2sss}
\end{figure}

\begin{lemma}\label{lem:sss-dual}
If we are given an $(a,b)$-CSP instance, we can find an equivalent
$(b,a)$-CSP instance in which each constraint of the $(a,b)$-CSP instance
corresponds to a single variable of the transformed problem,
and each constraint of the transformed problem corresponds to a single
variable of the original problem.
\end{lemma}

\begin{proof}
An assignment of colors to the original $(a,b)$-CSP instance's variables
solves the problem if and only if, for each constraint, there is at least
one pair $(V,C)$ in the constraint that does not appear in the coloring.
In our transformed problem, we choose one variable per original
constraint, with the colors available to the new variable being these
pairs $(V,C)$ in the corresponding constraint in the original problem.
Choosing such a pair in a coloring of the transformed problem is
interpreted as ruling out $C$ as a possible color for $V$ in the original
problem.  We then add constraints to our transformed problem to ensure
that for each
$V$ there remains at least one color that is not ruled out: we add one
constraint for each
$a$-tuple of colors of new variables---recall that each such color is a
pair $(V,C)$---such that all colors in the $a$-tuple
involve the same original variable $V$ and exhaust all the
choices of colors for $V$.\end{proof}

This duality may be easier to understand with a small example.
As discussed above, 3-SAT is essentially the same as $(2,3)$-CSP,
so Lemma~\ref{lem:sss-dual} can be used to translate 3-SAT to $(3,2)$-CSP.
Suppose we start with the 3-SAT instance
$(x_1\vee x_2\vee\neg{x}_3)\wedge(\neg{x}_1\vee x_3 \vee x_4)\wedge
(x_1\vee\neg{x}_2\vee\neg{x}_4)$.
Then we make a $(3,2)$-CSP instance (Figure~\ref{fig:3s2sss}) with three
variables
$v_i$, one for each 3-SAT clause.  Each variable has three possible
colors:
$(1,2,3)$ for $v_i$, $(1,3,4)$ for $v_2$,
and $(1,2,4)$ for $v_3$.
The requirement that value $T$ or $F$ be available to $x_1$
corresponds to the constraints $((v_1,1),(v_2,1))$
and $((v_2,1),(v_3,1))$; we similarly get constraints
$((v_1,2),(v_3,2))$, $((v_1,3),(v_2,3))$, and $((v_2,4),(v_3,4))$.
One possible coloring of this $(3,2)$-CSP instance would be
to color $v_1$ $1$, $v_2$ $3$, and $v_3$ $4$; this would give satisfying
assignments in which $x_1$ and $x_3$ are $T$, $x_4$ is $F$,
and $x_2$ can be either $T$ or $F$.

We can similarly translate an $(a,a)$-CSP instance into
an $(a,2)$-CSP instance in which each variable corresponds to either a
constraint or a variable, and each constraint forces the variable colorings
to match up with the dual constraint colorings; we omit the details as
we do not use this construction in our algorithms.

\begin{figure}[t]
$$\includegraphics[width=2.6in]{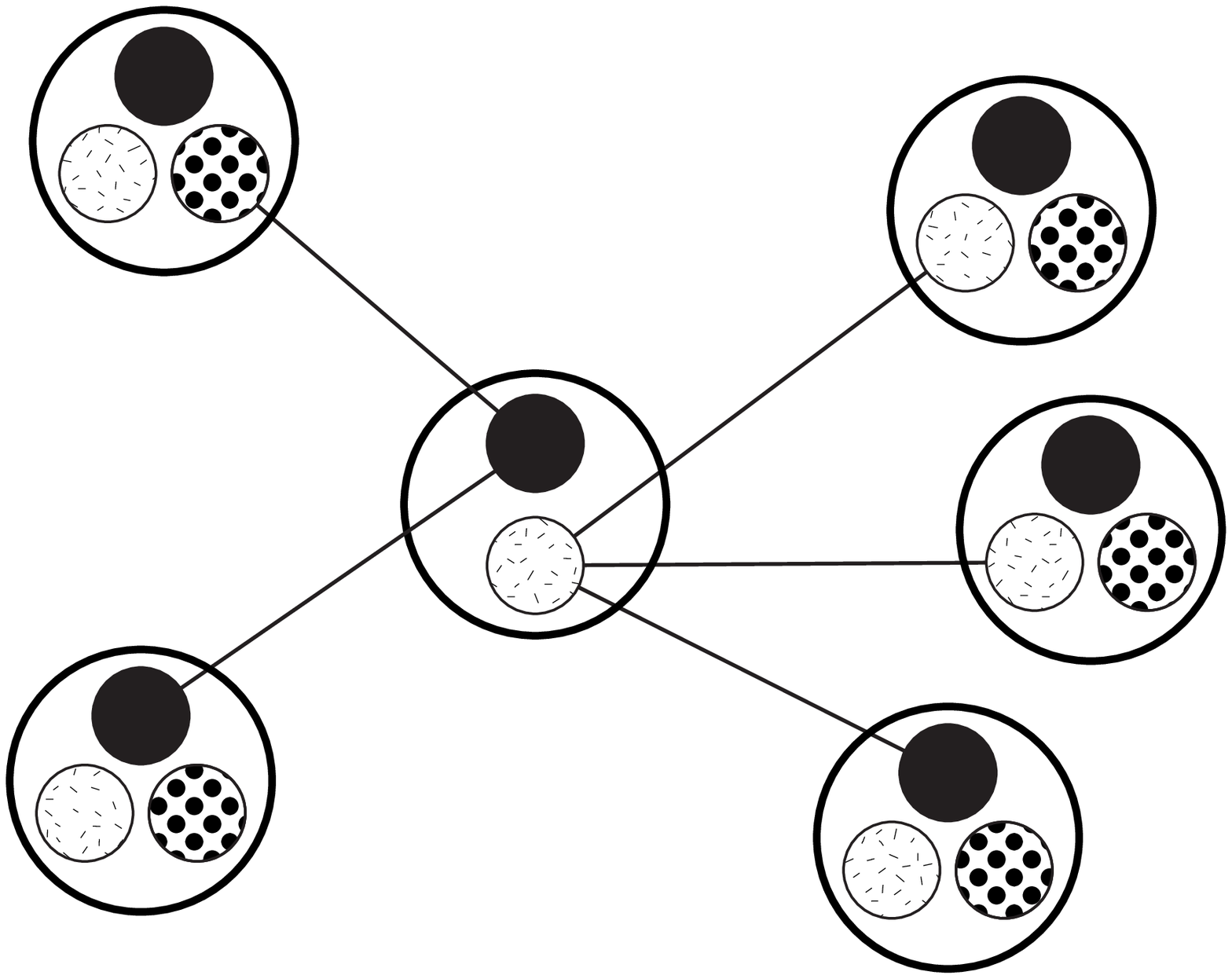}\qquad
\includegraphics[width=2.6in]{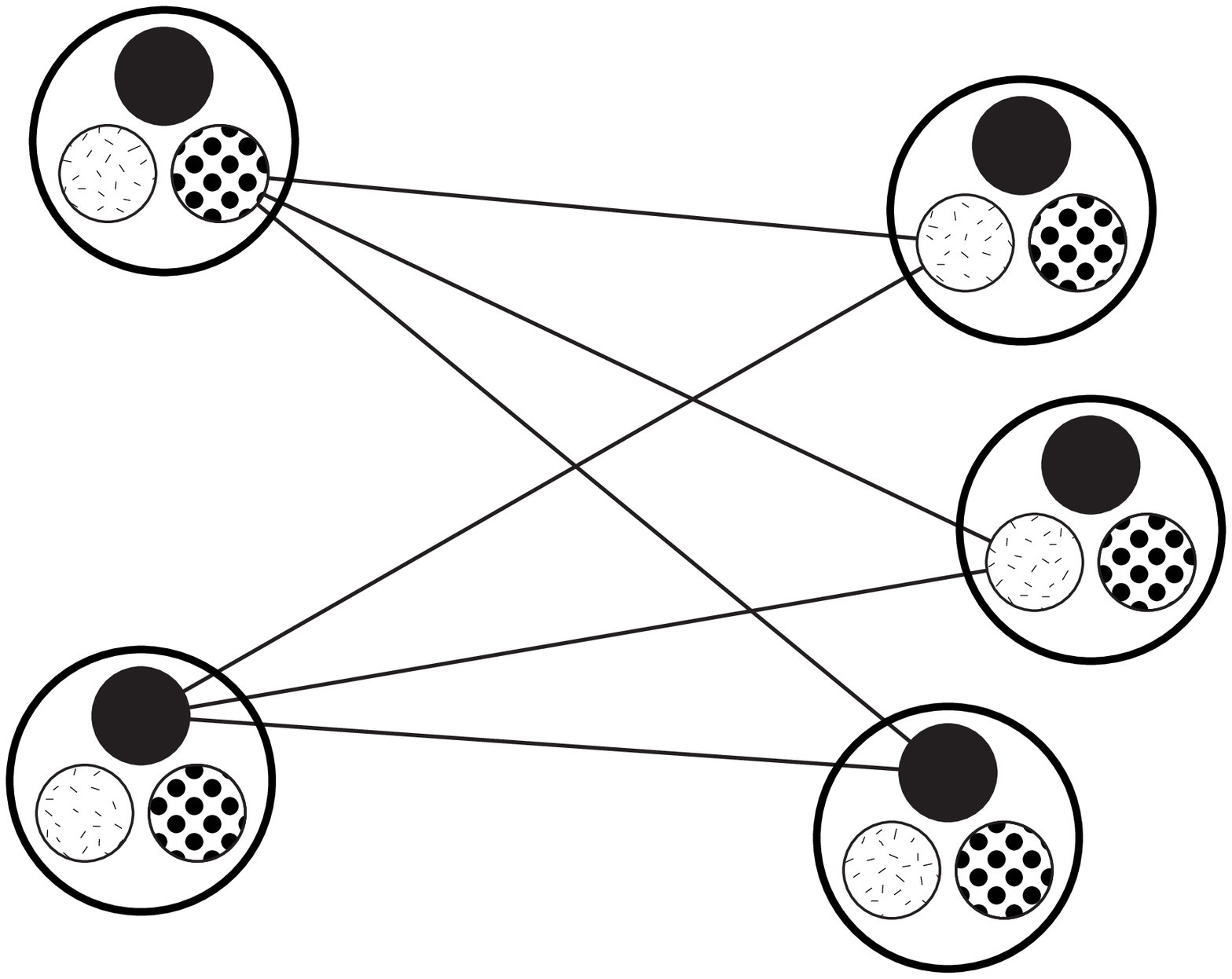}$$
\caption{$(3,2)$-CSP instance with a two-color variable (left)
and reduced instance after application of Lemma~\ref{lem:sss-elim}
(right).}
\label{fig:2elim}
\end{figure}

\section{Simplification of CSP Instances}

Before we describe our CSP algorithms, we describe some situations in
which the number of variables in an CSP instance may be reduced with
little computational effort.

\begin{lemma}\label{lem:sss-elim}
Let $v$ be a variable in an $(a,2)$-CSP instance,
such that only two of the $a$ colors are allowed at $v$.
Then we can find an equivalent $(a,2)$-CSP instance with one fewer
variable.
\end{lemma}

\begin{proof}
Let the two colors allowed at $v$ be $R$ and $G$.
Define ${\rm conflict}(C)$ to be the set of pairs
$\{(u,A):((u,A),(v,C))$ is a constraint.
We then include ${\rm conflict}(R)\times{\rm conflict}(G)$
to our set of constraints.

Any pair $((u,A),(w,B))\in{\rm conflict}(R)\times{\rm conflict}(G)$
does not reduce the space of solutions to the original problem
since if both $(u,A)$ and $(w,B)$ were present in a coloring
there would be no possible color left for $v$.
Conversely if all such constraints are satisfied,
one of the two colors for $v$ must be available.
Therefore we can now find a smaller equivalent problem by removing~$v$,
as shown in Figure~\ref{fig:2elim}.
\end{proof}

When we apply this variable elimination scheme, the number of
constraints can increase, but there can exist only $(an)^2$ distinct
constraints, which in our applications will be a small polynomial.

\begin{lemma}\label{lem:freesub}
Let $(v,X)$ and $(w,Y)$ be (variable,color) pairs in an $(a,2)$-CSP
instance, such that $v\ne w$ the only constraints involving these pairs
are either of the form $((v,X),(w,Z))$ with $Y\ne Z$,
or $((v,Z),(w,Y))$ with $X\ne Z$.
Then we can find an equivalent $(a,2)$-CSP instance with two fewer
variables.
\end{lemma}

\begin{proof}
It is safe to choose the colors $(v,X)$ and $(w,Y)$, since
these two choices do not conflict with each other nor with anything else
in the CSP instance.
\end{proof}

\begin{lemma}\label{lem:domsub}
Let $(v,R)$ and $(v,B)$ be (variable,color) pairs in an $(a,2)$-CSP
instance, such that whenever the instance contains a constraint
$((v,R),(w,X))$ it also contains a constraint $((v,B),(w,X))$.
Then we can find an equivalent $(a,2)$-CSP instance with one fewer
variable.
\end{lemma}

\begin{proof}
Any solution involving $(v,B)$ can be changed to one involving $(v,R)$
without violating any additional constraints, so it is safe to remove
the option of coloring $v$ with color $B$.  Once we remove this option,
$v$ is restricted to two colors, and we can apply
Lemma~\ref{lem:sss-elim}.
\end{proof}

\begin{lemma}\label{lem:goodsub}
Let $(v,R)$ be a (variable,color) pair in an $(a,b)$-CSP instance
that is not involved in any constraints.
Then we can find an equivalent $(a,b)$-CSP instance with one fewer
variable.
\end{lemma}

\begin{proof}
We may safely assign color $R$ to $v$ and remove it from the instance.
\end{proof}

\begin{lemma}\label{lem:badsub}
Let $(v,R)$ be a (variable,color) pair in an $(a,2)$-CSP instance
that is involved in constraints with all three color options of another
variable $w$. Then we can find an equivalent $(a,b)$-CSP instance with one
fewer variable.
\end{lemma}

\begin{proof}
No coloring of the instance can use $(v,R)$, so we can restrict $v$ to
the remaining two colors and apply Lemma~\ref{lem:sss-elim}.
\end{proof}

We say that a CSP instance in which none of Lemmas
\ref{lem:sss-elim}--\ref{lem:badsub} applies is {\em reduced}.

\section{Simple Randomized CSP Algorithm}

\begin{figure}[t]
$$\includegraphics[width=5in]{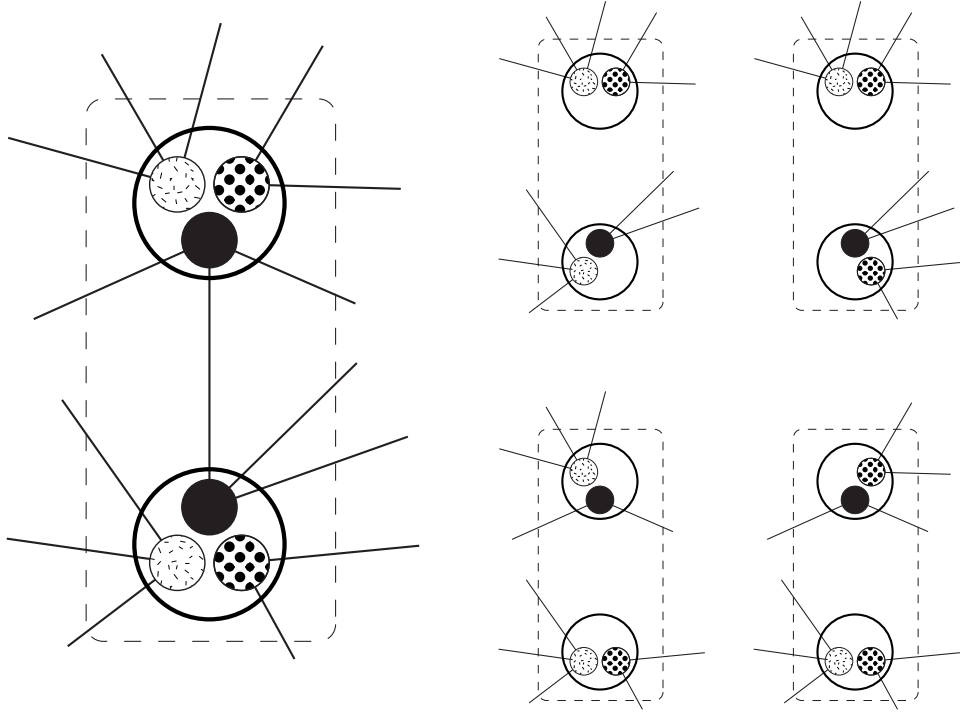}$$
\caption{Randomized $(3,2)$-CSP algorithm: constraint $((v,R),(w,R))$
(left) and four equally-likely two-color restrictions in which exactly
one of the variables may still use color $R$ (right).}
\label{fig:random}
\end{figure}

We first demonstrate the usefulness of Lemma~\ref{lem:sss-elim}
by describing a very simple randomized algorithm for solving
$(3,2)$-CSP instances in expected time $O(2^{n/2}n^{O(1)})$.

\begin{lemma}
If we are given a $(3,2)$-CSP instance $I$, then in random polynomial time
we can find an instance $I'$ with two fewer variables, such that
if $I'$ is solvable then so is $I$, and if $I$ is solvable then with
probability at least $\frac12$ so is $I'$.
\end{lemma}

\begin{proof}
If no constraint exists, we can solve the problem immediately.
Otherwise choose some constraint $((v,X),(w,Y))$.
Rename the colors if necessary so that both $v$ and $w$
have available the same three colors $R$, $G$, and $B$,
and so that $X=Y=R$.
Restrict the colorings of $v$ and $w$ to two colors each
in one of four ways, chosen uniformly at random from the four possible
such restrictions in which exactly one of $v$ and $w$
is restricted to colors $G$ and $B$ (Figure~\ref{fig:random}).
Then it can be verified by examination of cases that any valid coloring of
the problem remains valid for exactly two of these four restrictions,
so with probability
$\frac12$ it continues to be a solution to the restricted problem.
Now apply Lemma~\ref{lem:sss-elim} and eliminate both $v$ and $w$ from the
problem.
\end{proof}

\begin{corollary}
In expected time $O(2^{n/2}n^{O(1)})$ we can find a solution to a
$(3,2)$-CSP instance if one exists.
\end{corollary}

\begin{proof}
We perform the reduction above $n/2$ times, taking polynomial time
and giving probability at least $2^{-n/2}$ of finding a correct solution.
If we repeat this method until a solution is found, the expected
number of repetitions is $2^{n/2}$.
\end{proof}

\section{Faster CSP Algorithm}

We now describe a more complicated method of solving $(3,2)$-CSP
instances deterministically with the somewhat
better time bound of $O(1.36443^n)$.
More generally, our algorithm can actually handle $(4,2)$-CSP instances.
Any $(4,2)$-CSP instance can be transformed into a $(3,2)$-CSP instance
by expanding each of its four-color variables to two three-color
variables, each having two of the original four colors, with a constraint
connecting the third color of each new variable (Figure~\ref{fig:iso33}).
Therefore, the natural definition of the ``size'' of a $(4,2)$-CSP
instance is
$n=n_3+2n_4$, where $n_i$ denotes the number of variables with $i$ colors.
However, we instead define the size to be $n=n_3+(2-\epsilon)n_4$,
where $\epsilon\approx 0.095543$ is a constant to be determined more
precisely later.  In any case, the size of a $(3,2)$-CSP instance
remains equal to its number of variables, so any bound on the running
time of our algorithm in terms of $n$ applies directly to $(3,2)$-CSP.

The basic idea of our algorithm is to find a set of local configurations
that must occur within any $(4,2)$-CSP instance $I$, such that any
instance containing such a configuration can be replaced by a small
number of smaller instances.

\def\rec#1{\lambda(#1)}

In more detail,
for each configuration we describe a
set of smaller instances $I_i$ of size $|I|-r_i$ such that $I$ is solvable
if and only if at least one of the instances $I_i$ is solvable.  If one
particular configuration occurred at each step of the algorithm, this
would lead to a recurrence of the form
$$T(n)=\sum T(n-r_i)+\hbox{poly}(n)=O(\rec{r_1,r_2,\ldots}^n)$$
for the worst-case running time of our algorithm, where the base
$\rec{r_1,r_2,\ldots}$ of the exponent in the running time is the largest
zero of the function
$f(x)=1-\sum x^{-r_i}$ (such a function is not necessarily a polynomial
because the $r_i$ will not necessarily be integers).
We call this value $\rec{r_1,r_2,\ldots}$
the {\em work factor} of the given local configuration.
The overall time bound will be $\lambda^n$ where $\lambda$ is the largest
work factor among the configurations we have identified.
This value $\lambda$ will depend on our previous choice of $\epsilon$; we
will choose $\epsilon$ in such a way as to minimize $\lambda$.

\subsection{Single Constraints and Multiple Adjacencies}

\begin{figure}[t]
$$\includegraphics[width=4in]{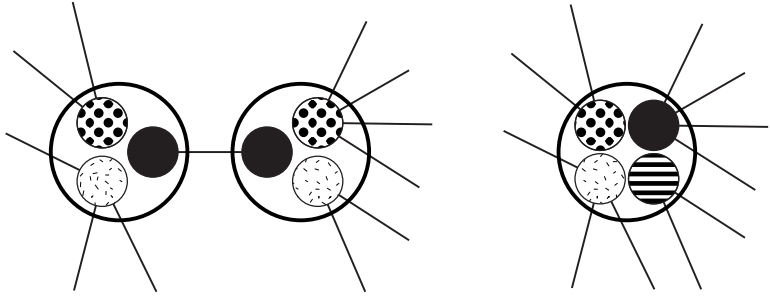}$$
\caption{Isolated constraint between two three-color variables (left) can
be replaced by a single four-color variable (right).}
\label{fig:iso33}
\end{figure}

\begin{figure}[t]
$$\includegraphics[width=4in]{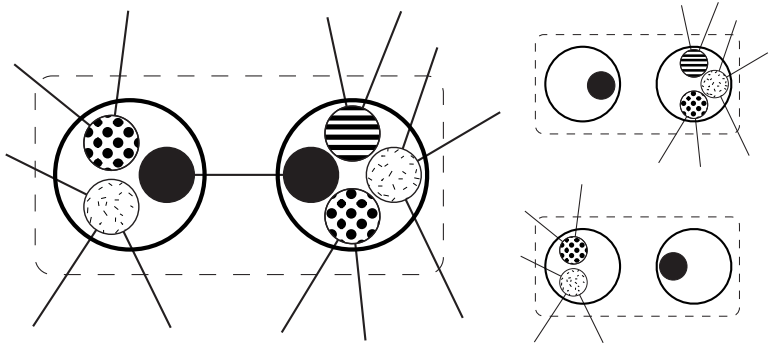}$$
\caption{Isolated constraint between three-color and four-color variable
(left) can be replaced by two instances with size smaller by $2-\epsilon$
(right top) and $3-\epsilon$ (right bottom).}
\label{fig:iso34}
\end{figure}

We first consider local configurations in which some (variable,color)
pair is incident on only one constraint, or has multiple constraints to
the same variable.
First, suppose that (variable,color) pair
$(v,R)$ is involved in only a single constraint $((v,R),(w,R))$.
If this is also the only constraint involving $(w,R)$, we call it an
{\em isolated constraint}.  Otherwise, we call it a {\em dangling
constraint}.

\begin{lemma}\label{lem:isolated}
Let $((v,R),(w,R))$ be an isolated constraint in a $(4,2)$-CSP instance,
and let $\epsilon\le 0.545$.
Then the instance can be replaced by smaller instances
with work factor at most $\rec{2-\epsilon,3-\epsilon}$.
\end{lemma}

\begin{proof}
If $v$ and $w$ are both three-color variables, then the instance can be
colored if and only if we can color the instance formed by replacing them
with a single four-color variable, in which the four colors are the
remaining choices for $v$ and $w$ other than $R$
(Figure~\ref{fig:iso33}).  Thus in this case we can reduce the problem
size by $\epsilon$, with no additional work.

Otherwise, if there exists a coloring of the given instance, there
exists one in which exactly one of $v$ and $w$ is given color $R$.
Suppose first that $v$ has four colors while $w$ has only three.
Thus we can reduce the problem to two instances, in one of which $(v,R)$
is used (so $v$ is removed from the problem, and $(w,R)$ is removed as a
choice for variable $w$, allowing us to remove the variable by
Lemma~\ref{lem:sss-elim}) and in the other of which
$(w,R)$ is used (Figure~\ref{fig:iso34}). The first subproblem has its
size reduced by
$3-\epsilon$ since both variables are removed,
while the second's size is reduced by $2-\epsilon$ since $w$ is removed
while $v$ loses one of its colors but is not removed.
Thus the work factor is $\rec{2-\epsilon,3-\epsilon}$. Similarly, if
both are four-color variables, the work factor is
$\rec{3-2\epsilon,3-2\epsilon}$.
For the given range of $\epsilon$, this second work factor is smaller
than the first.
\end{proof}

\begin{figure}[t]
$$\includegraphics[width=5in]{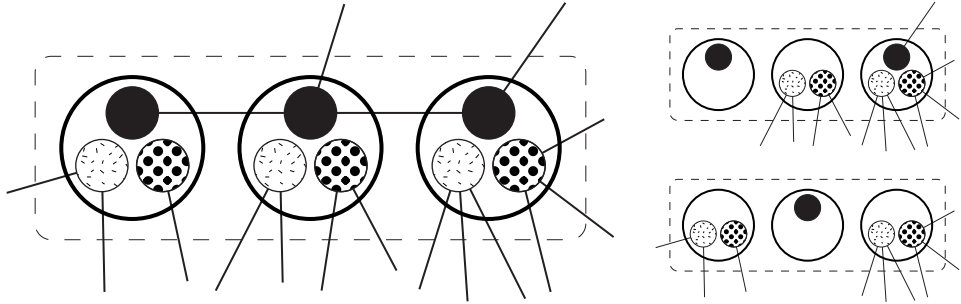}$$
\caption{Dangling constraint: a choice with one constraint,
where the second choice in the constraint is also constrained by a third
variable. We choose either to use or not use that second color.}
\label{fig:dangle}
\end{figure}

\begin{lemma}\label{lem:dangle}
Let $((v,R),(w,R))$ be a dangling constraint in a reduced $(4,2)$-CSP
instance.
Then the instance can be replaced by smaller instances
with work factor at most $\rec{2-\epsilon,3-\epsilon}$.
\end{lemma}

\begin{proof}
The second constraint for $(w,R)$ can not involve $v$,
or we would be able to apply Lemma~\ref{lem:domsub}.
We choose either to use color $(w,R)$ or to restrict $w$ to avoid that
color (Figure~\ref{fig:dangle}).  If we use color $(w,R)$, we
eliminate choice $(v,R)$ and another choice on the other neighbor of
$w$. If we avoid color $(w,R)$, we may safely use color
$(v,R)$.

In the worst case, the other neighbor of $(w,R)$ has four colors, so
removing one only reduces the problem size by $1-\epsilon$.  There are
four cases depending on the number of colors of $v$ and $w$:
If both have three colors, the work factor is $\rec{2,3-\epsilon}$.
If only $v$ has four colors, the work factor is
$\rec{3-\epsilon,3-2\epsilon}$.
If only $w$ has four colors, the work factor is
$\rec{2-\epsilon,4-2\epsilon}$.
If both have four colors, the work factor is
$\rec{3-2\epsilon,4-3\epsilon}$.
These factors are all dominated by the one in the statement of the lemma.
\end{proof}

\begin{figure}[t]
$$\includegraphics[width=4.25in]{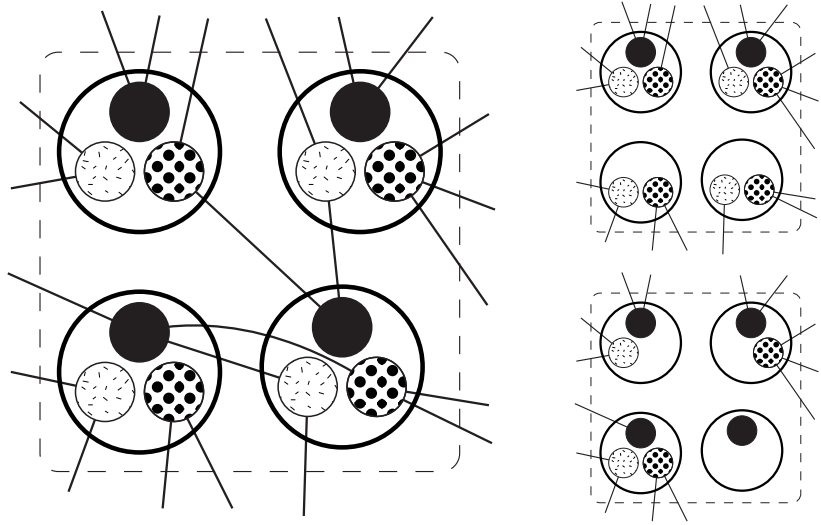}$$
\caption{Implication from $(v,R)$ to $(w,R)$, such that
$(w,R)$ has two distinct neighbors.
Restricting $w$ eliminates $v$ and $w$ (top right)
while assigning $w$ color $R$ eliminates three variables (bottom right).}
\label{fig:multadj}
\end{figure}

\begin{figure}[t]
$$\includegraphics[width=2.5in]{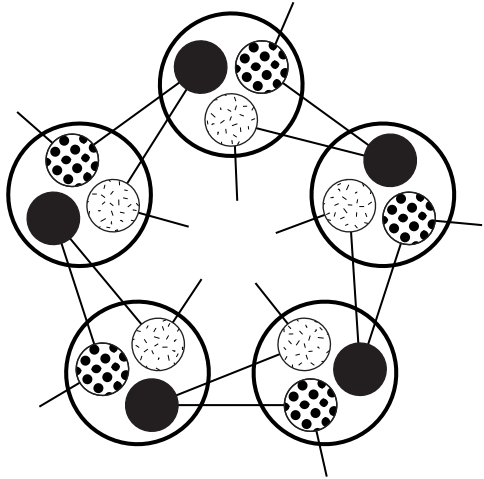}$$
\caption{Cycle of implications.}
\label{fig:macycle}
\end{figure}

\begin{lemma}\label{lem:multadj}
Suppose a reduced $(4,2)$-CSP instance includes
two constraints such as $((v,R),(w,B))$ and $((v,R),(w,G))$ that connect
one color of variable $v$ with two colors of variable $w$,
and let $\epsilon\le 0.4$. Then the
instance can be replaced by smaller instances with work factor at most
$\rec{2-\epsilon,3-2\epsilon}$.
\end{lemma}

\begin{proof}
We assume that the instance has no color choice with only a single
constraint, or we could apply one of Lemmas \ref{lem:isolated}
and~\ref{lem:dangle} to achieve the given work factor.

We say that $(v,R)$ implies $(w,R)$ if there
are constraints from
$(v,R)$ to every other color choice of $w$.  If the target $(w,R)$ of an
implication is not the source of another implication, then using
$(w,R)$ eliminates $w$ and at least two other colors, while avoiding
$(w,R)$ forces us to also avoid $(v,R)$ (Figure~\ref{fig:multadj}). Thus,
in this case we achieve work factor either $\rec{2-\epsilon,3-2\epsilon}$
if $w$ has three color choices, or $\rec{2-2\epsilon,4-3\epsilon}$ if it
has four.

If the target of every implication is the source of another, then we can
find a cycle of colors each of which implies the next in the cycle
(Figure~\ref{fig:macycle}). If no other constraints involve
colors in the cycle (as is true in the figure), we can use them all,
reducing the problem by the length of the cycle for free. Otherwise, let
$(v,R)$ be a color in the cycle that has an outside constraint.  If we use
$(v,R)$, we must use the colors in the rest of the cycle, and eliminate
the (variable,color) pair outside the cycle constrained by
$(v,R)$.
If we avoid $(v,R)$, we must also avoid the colors in the rest of the
cycle.
The maximum work factor for this case is $\rec{2,3-\epsilon}$,
and arises when the cycle consists of only two variables,
both of which have only three allowed colors.

Finally, if the situation described in the lemma exists without forming
any implication, then $w$ must have four color choices, exactly two of
which are constrained by $(v,R)$.  In this case restricting $w$ to those
two choices reduces the size by at least $3-2\epsilon$, while
restricting it to the remaining two choices reduces the size by
$2-\epsilon$, again giving work factor $\rec{2-\epsilon,3-2\epsilon}$.
\end{proof}

\subsection{Highly Constrained Colors}

We next consider cases in which choosing one color for a variable
eliminates many other choices, or in which
adjacent (variable,color) pairs have different numbers of constraints.

\begin{lemma}\label{lem:highdeg}
Suppose a reduced $(4,2)$-CSP instance includes a color
pair $(v,R)$ involved in three or more constraints,
where $v$ has four color choices,
or a pair $(v,R)$ involved in four or more constraints,
where $v$ has three color choices.
Then the
instance can be replaced by smaller instances with work factor at most
$\rec{1-\epsilon,5-4\epsilon}$.
\end{lemma}

\begin{proof}
We can assume from Lemma~\ref{lem:multadj} that each constraint connects
$(v,R)$ to a different variable.  Then if we choose to use color
$(v,R)$, we eliminate $v$ and remove a choice from each of its
neighbors, either eliminating them or reducing their number of choices
from four to three.  If we don't use $(v,R)$, we eliminate that color
only.  So if $v$ has four choices, the work factor
is at most $\rec{1-\epsilon,5-4\epsilon}$,
and if it has three choices and four or more constraints,
the work factor is at most
$\rec{1,5-4\epsilon}$.
\end{proof}

\begin{figure}[t]
$$\includegraphics[width=5in]{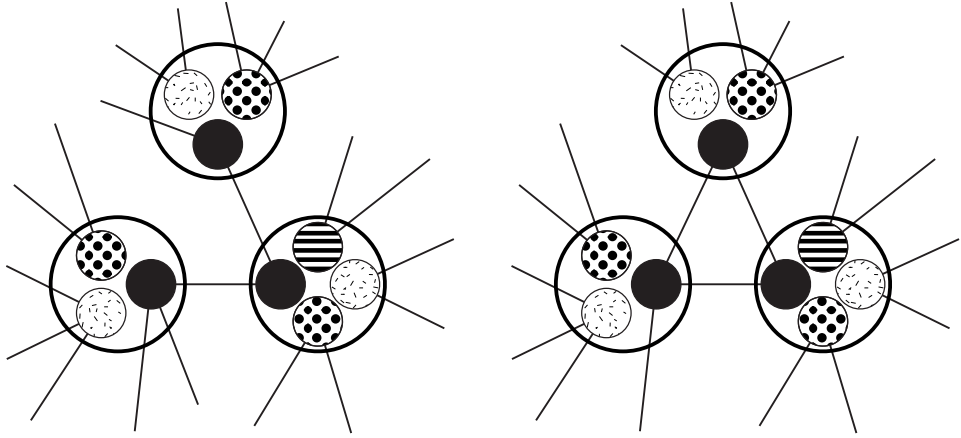}$$
\caption{Cases for Lemma~\ref{lem:42-highdeg}:
$(v,R)$ and $(w,R)$ either have disjoint neighbors (left) or form a
triangle (right).}
\label{fig:3con4ch}
\end{figure}

\begin{lemma}\label{lem:42-highdeg}
Suppose a reduced $(4,2)$-CSP instance includes a (variable,color)
pair $(v,R)$ with three constraints,
one of which connects it to a variable with four color choices,
and let $\epsilon\le 0.3576$.
Suppose also that none of the previous lemmas applies.
Then the
instance can be replaced by smaller instances with work factor at most
$\rec{3-\epsilon,4-\epsilon,4-\epsilon}$.
\end{lemma}

\begin{proof}
For convenience suppose that the four-color neighbor is $(w,R)$.
We can assume $(w,R)$ has only two constraints, else it would be covered
by a previous lemma.

Then, if $(v,R)$ and $(w,R)$ do not form a triangle with a third
(variable,color) pair (Figure~\ref{fig:3con4ch}, left), we choose either
to use or avoid color
$(v,R)$. If we use
$(v,R)$, we eliminate $v$ and the three adjacent color choices. If we
avoid $(v,R)$, we create a dangling constraint at $(w,R)$, which we have
seen in Lemma~\ref{lem:dangle} allows us to further subdivide the instance
with work factor $\rec{3-\epsilon,3-2\epsilon}$ in addition to the
elimination of
$v$. Thus, the overall work factor in this case is
$\rec{4-\epsilon,4-2\epsilon,4-3\epsilon}$.

On the other hand, suppose we have a triangle of constraints formed by
$(v,R)$, $(w,R)$, and a third (variable,color) pair $(x,R)$, as
shown in Figure~\ref{fig:3con4ch}, right.
Then $(v,R)$ and $(x,R)$ are the only choices constraining $(w,R)$,
so if $(v,R)$ and $(x,R)$ are both not chosen, we can safely choose to
use color $(w,R)$.  Therefore, we make three smaller instances, in each
of which we choose to use one of the three choices in the triangle.
We can assume from the previous cases that $(v,R)$ has only three
choices, and further its third neighbor (other than $(w,R)$ and $(x,R)$)
must also have only three choices or we could apply the previous case of
the lemma.  In the worst case, $(x,R)$ has only two constraints and $x$
has only three color choices.  Therefore, the size of the subproblems
formed by choosing $(v,R)$, $(w,R)$, and $(x,R)$ is
reduced by at least $4-\epsilon$, $4-\epsilon$, and $3-\epsilon$
respectively, leading to a work factor of
$\rec{3-\epsilon,4-\epsilon,4-\epsilon}$.  If instead $x$ has four color
choices, we get the better work factor
$\rec{4-2\epsilon,4-2\epsilon,4-2\epsilon}$.

For the given range of $\epsilon$, the largest of these work factors is
$\rec{3-\epsilon,4-\epsilon,4-\epsilon}$.
\end{proof}

\begin{figure}[t]
$$\includegraphics[width=5in]{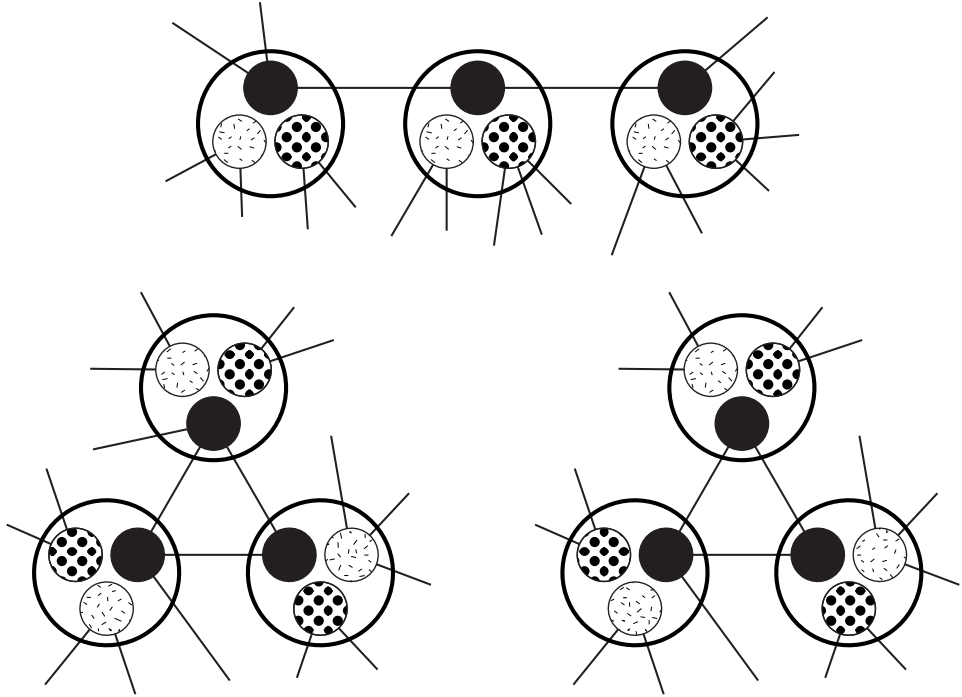}$$
\caption{Cases for Lemma~\ref{lem:32-highdeg}.
Top: $(v,R)$ and $(w,R)$ do not form a triangle;
avoiding $(v,R)$ produces a dangling constraint.
Bottom left: $(v,R)$ and $(w,R)$ are part of a triangle
with two triply-constrained colors;
choosing each triangle vertex gives work factor
$\rec{3,4,4}$.
Bottom right: $(v,R)$ and $(w,R)$ are part of a triangle
with two doubly-constrained colors;
avoiding $(v,R)$ produces an isolated constraint.}
\label{fig:3con2con}
\end{figure}

\begin{lemma}\label{lem:32-highdeg}
Suppose a reduced $(4,2)$-CSP instance includes a (variable,color)
pair $(v,R)$ with three constraints,
one of which connects it to a variable with two constraints.
Suppose also that none of the previous lemmas applies.
Then the
instance can be replaced by smaller instances with work factor at most
$\max\{\rec{1+\epsilon,4},\rec{3,4-\epsilon,4}\}$.
\end{lemma}

\begin{proof}
Let $(w,R)$ be the neighbor with two constraints.
Note that (since the previous lemma is assumed not to apply) all
neighbors of $(v,R)$ have only three color choices.

First, suppose $(v,R)$ and $(w,R)$ are not part of a triangle of
constraints (Figure~\ref{fig:3con2con}, top). Then, if we choose to use
color
$(v,R)$ we eliminate four variables, while if we avoid using it we create
a dangling constraint on
$(w,R)$ which we further subdivide into two more instances according to
Lemma~\ref{lem:dangle}.  Thus, the work factor in this case is
$\rec{3,4-\epsilon,4}$.

Second, suppose that $(v,R)$ and $(w,R)$ are part of a triangle with a
third (variable,color) pair $(x,R)$, and that $(x,R)$ has three
constraints (Figure~\ref{fig:3con2con}, bottom left).  Then (as in the
previous lemma) we may choose to use one of the three choices in the
triangle, resulting in work factor $\rec{3,4,4}$.

Finally, suppose that $(v,R)$, $(w,R)$, and $(x,R)$ form a triangle as
above, but that $(x,R)$ has only two constraints
(Figure~\ref{fig:3con2con}, bottom right).  Then if we choose to use
$(v,R)$ we eliminate four variables, while if we avoid using it we create
an isolated constraint between $(w,R)$ and $(x,R)$. Thus in this case the
work factor is $\rec{1+\epsilon,4}$.
\end{proof}

If none of the above lemmas applies to an instance, then each color choice
in the instance must have either two or three constraints, and each
neighbor of that choice must have the same number of constraints.

\subsection{Triply-Constrained Colors}

\begin{figure}[t]
$$\includegraphics[width=3.5in]{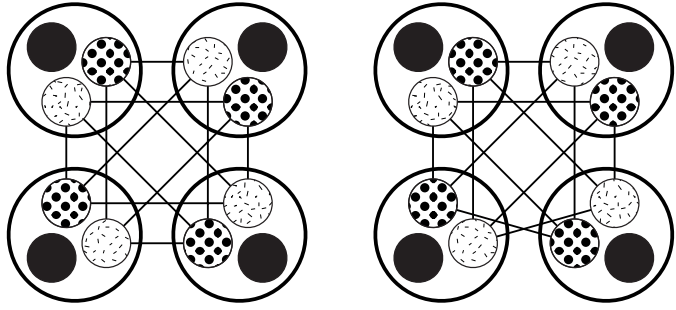}$$
\caption{The two possible small three-components with $k=8$.}
\label{fig:small3}
\end{figure}

Within this section we assume that we have a $(4,2)$-CSP instance in
which none of the previous reduction lemmas applies, so any
(variable,color) pair must be involved in exactly as many constraints
as each of its neighbors.

We now consider
the remaining (variable,color) pairs that have three constraints each. 
Define a {\em three-component} to be a subset of such pairs such that any
pair in the subset is connected to any other by a path of constraints.  We
distinguish two such types of components: a {\em small three-component} is
one that involves only four distinct variables, while a {\em large
three-component} involves five or more variables.  Note that we can
assume by the previous lemmas that each variable in a component has only
three color choices.

\begin{lemma}
Let $C$ be a small three-component involving $k$ (variable,color) pairs.
Then $k$ must be a multiple of four, and each variable involved in the
component has exactly $k/4$ pairs in $C$.
\end{lemma}

\begin{proof}
Let $v$ and $w$ be variables in a small component $C$.  Then each
(variable,color) pair in $C$ from variable $v$ has exactly one
constraint to a distinct (variable,color) pair from variable $w$,
so the numbers of pairs from $v$ equals the number of pairs from $w$.
The assertions that each variable has the same number of pairs,
and that the total number of pairs is a multiple of four, then follow.
\end{proof}

We say that a small three-component is {\em good} if $k=4$ in the lemma
above.

\begin{lemma}
Let $C$ be a small three-component that is not good.
Then the
instance can be replaced by smaller instances with work factor at most
$\rec{4,4,4}$.
\end{lemma}

\begin{proof}
A component with $k=12$ uses up all color choices for all four
variables.  Thus we may consider these variables in isolation from the
rest of the instance, and either color them all (if possible) or
determine that the instance is unsolvable.

The remaining small components have $k=8$.
Such a component may be drawn with the four variables at the corners of
a square, and the top, left, and right pairs of edges uncrossed
(Figure~\ref{fig:small3}).
If only the center two pairs were crossed, we would actually have two
$k=4$ components, and if any other two or three of the remaining pairs
were crossed, we could reduce the number of crossings in the drawing by
swapping the colors at one of the variables.  Thus, the only possible
small components with $k=8$ are the one with all six pairs uncrossed, and
the one with only one pair crossed.

The first of these allows all four variables to be colored and removed,
while in the other case there exist only three maximal subsets of
variables that can be colored. (In the figure, these three sets are
formed by the bottom two vertices, and the two sets formed by removing
one bottom vertex).  We split into instances by choosing to color each of
these maximal subsets, eliminating all four variables in the component and
giving work factor $\rec{4,4,4}$.
\end{proof}

\begin{figure}[t]
$$\includegraphics[width=4.75in]{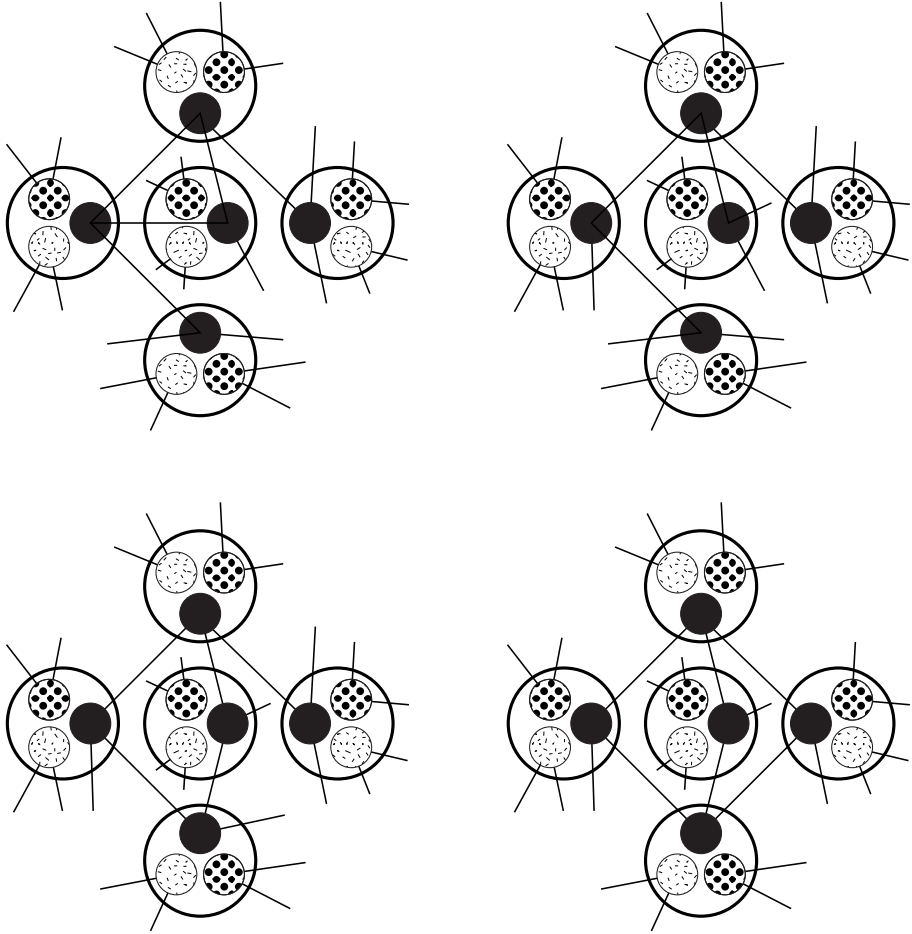}$$
\caption{Cases for Lemma~\ref{lem:large3}:
$(z,R)$ is constrained by one neighbor of $(v,R)$ forming a triangle
with a second neighbor (top left), one neighbor not part of a triangle
(top right), two neighbors (bottom left), or three neighbors (bottom
right).}
\label{fig:witness}
\end{figure}

Define a {\em witness} to a large three-component to be a set of five
(variable,color) pairs with five distinct variables, such that there
exist constraints from one pair to three others, and from at least one
of those three to the fifth.  By convention we use $(v,R)$ to denote the
first pair, $(w,R)$, $(x,R)$, and $(y,R)$ to denote the pairs connected
by constraints to $(v,R)$, and $(z,R)$ to be the fifth pair in the
witness.

\begin{lemma}\label{lem:witness}
Every large three-component has a witness.
\end{lemma}

\begin{proof}
Choose some arbitrary pair $(u,R)$ as a starting point, and perform a
breadth first search in the graph formed by the pairs and constraints in
the component.  Let $(z,R)$ be the first pair reached by this search
where $z$ is not one of the variables adjacent to $(u,R)$,
let $(v,R)$ be the grandparent of $(z,R)$ in the breadth first search
tree, and let the other three pairs be the neighbors of $(v,R)$.  Then
it is easy to see that $(v,R)$ and its neighbors must use the same four
variables as $(u,R)$ and its neighbors, while $z$ by definition uses a
different variable.
\end{proof}

\begin{lemma}\label{lem:large3}
Suppose that a
$(4,2)$-CSP instance contains a large three-component.
Then the
instance can be replaced by smaller instances with work factor at most
$\rec{4,4,5,5}$.
\end{lemma}

\begin{proof}
Let $(v,R)$, $(w,R)$, $(x,R)$, $(y,R)$, and $(z,R)$ be a witness for the
component.  Then we distinguish subcases according to how many of the
neighbors of $(z,R)$ are pairs in the witness.

\begin{enumerate}
\item If $(z,R)$ has a constraint with only one pair in the witness, say
$(w,R)$, then we choose either to use
color
$(z,R)$ or to avoid it. If we use it, we eliminate some four variables.
If we avoid it, then we cause $(w,R)$ to have only two constraints.
If $(w,R)$ is also constrained by one of $(x,R)$ or $(y,R)$,
we then have a triangle of constraints (Figure~\ref{fig:witness},
top left). We can assume without loss of generality that the remaining
constraint from this triangle does not connect to a different color of
variable $z$, for if it did we could instead use the same five variables
in a different order to get a witness of this form.
We then further subdivide into three more instances, in each of
which we choose to use one of the pairs in the triangle, as in the second
case of Lemma~\ref{lem:32-highdeg}. This gives overall work factor
$\rec{4,4,5,5}$.

On the other hand, if $(v,R)$ and $(w,R)$ are not part
of a triangle (Figure~\ref{fig:witness},
top right), then
(after avoiding $(z,R)$) we can apply the first case of
Lemma~\ref{lem:32-highdeg} again achieving the same work factor.

\item If $(z,R)$ has constraints with two pairs in the witness (Figure~\ref{fig:witness},
bottom left),
then choosing to use $(z,R)$ eliminates four variables and causes
$(v,R)$ to dangle, while avoiding $(z,R)$ eliminates a single variable.
The work factor is thus $\rec{1,6,7}$.

\item If $(z,R)$ has constraints with all three of $(w,R)$, $(y,R)$, and
$(z,R)$ (Figure~\ref{fig:witness},
bottom right), then choosing to use $(z,R)$ also allows us to use $(v,R)$,
eliminating five variables. The work factor is $\rec{1,5}$.
\end{enumerate}

\noindent
The largest of the three work factors arising in these cases is
the first one, $\rec{4,4,5,5}$.
\end{proof}

\subsection{Doubly-Constrained Colors}

As in the previous section, we define a {\em two-component} to be a subset
of (variable,color) pairs such that each has two constraints, and any pair
in the subset is connected to any other by a path of constraints.
A two-component must have the form of a cycle of pairs, but it is
possible for more than one pair in the cycle to involve the same
variable.  We distinguish two such types of components: a {\em small
two-component} is one that involves only three pairs, while
a {\em large two-component} involves four or more pairs.

\begin{lemma}\label{lem:cyclic}
Suppose a reduced $(4,2)$-CSP instance includes a large two-component,
and let $\epsilon\le 0.287$.
Then the instance can be replaced by smaller
instances with work factor at most $\rec{3,3,5}$.
\end{lemma}

\begin{proof}
We split into subcases:
\begin{enumerate}
\item Suppose the cycle passes through five consecutive distinct
variables, say $(v,R)$, $(w,R)$, $(x,R)$, $(y,R)$, and $(z,R)$.
We can assume that, if any of these five variables has four color
choices, then this is true of one of the first four variables.
Any coloring that does not use both $(v,R)$ and $(y,R)$
can be made to use at least one of the two colors $(w,R)$ or $(x,R)$
without violating any of the constraints.  Therefore, we can divide into
three subproblems: one in which we use $(w,R)$, eliminating three
variables, one in which  we use $(x,R)$, again eliminating three
variables, and one in which we use both $(v,R)$ and $(y,R)$,
eliminating all five variables.  If all five variables have
only three color choices,
The work factor resulting from this subdivision is $\rec{3,3,5}$.
If some of the variables have four color choices, the work factor is
at most $\rec{3-\epsilon,4-\epsilon,5-2\epsilon}$,
which is smaller for the given range of $\epsilon$.

\item Suppose two colors three constraints apart on a cycle belong to
the same variable; for instance, the sequence of colors may be
$(v,R)$, $(w,R)$, $(x,R)$, $(v,G)$.  Then any coloring can be made to
use one of $(w,R)$ or $(x,R)$ without violating any constraints.
If we form one subproblem in which we use $(w,R)$ and one in which we
use $(x,R)$, we get work factor at most $\rec{3-\epsilon,3-\epsilon}$
(the worst case occurring when only $v$ has four color choices).

\item Any long cycle which does not contain one of the previous two
subcases must pass through the same four variables in the same order one,
two, or three times.  If it passes through two or three times, all four
variables may be safely colored using colors from the cycle, reducing
the problem with work factor one.  And if the cycle has length exactly
four, we may choose one of two ways to use two diagonally opposite
colors from the cycle, giving work factor at most $\rec{4,4}$.
\end{enumerate}

\noindent
For the given range of $\epsilon$, the largest of these work factors is
$\rec{3,3,5}$.
\end{proof}

\subsection{Matching}

Suppose we have a $(4,2)$-CSP instance to which none of the preceding
reduction lemmas applies.
Then, every constraint must be part of a good three-component or a small
two-component.  As we now show, this simple structure enables us to
solve the remaining problem quickly.

\begin{lemma}\label{lem:matching}
If we are given a $(4,2)$-CSP instance in which every constraint must be
part of a good three-component or a small two-component,
then we can solve it or determine that it is not solvable in polynomial
time.
\end{lemma}

\begin{proof}
We form a bipartite graph, in which the vertices correspond to the
variables and components of the instance.  We connect a variable to a
component by an edge if there is a (variable,color) pair using that
variable and belonging to that component.

Since each pair in a good three-component or small two-component is
connected by a constraint to every other pair in the component,
any solution to the instance can use at most one (variable,color) pair
per component.  Thus, a solution consists of a set of (variable,color)
pairs, covering each variable once, and covering each component at most
once.  In terms of the bipartite graph constructed above, this is simply
a matching.  So, we can solve the problem by using a graph
maximum matching algorithm to determine the existence of a matching that
covers all the variables.
\end{proof}

\subsection{Overall CSP Algorithm}

This completes the case analysis needed for our result.

\begin{theorem}\label{thm:sss}
We can solve any $(3,2)$-CSP instance in time 
$O(\rec{4,4,5,5}^n)\approx O(1.36443^n)$.
\end{theorem}

\begin{proof}
We employ a backtracking (depth first) search in a state space consisting
of $(3,2)$-CSP instances.  At each point in the search, we examine the
current state, and attempt to find a set of smaller instances to replace
it with, using one of the reduction lemmas above.
Such a replacement can always be found in
polynomial time by searching for various simple local configurations in
the instance.  We then recursively search each smaller instance in
succession.  If we ever reach an instance in which
Lemma~\ref{lem:matching} applies, we perform a matching algorithm to test
whether it is solvable. If so, we find a solution and terminate the
search.  If not, we backtrack to the most recent branching point of the
search and continue with the next alternative at that point.

A bound of $\lambda^n$ on the number of recursive calls in this
search algorithm, where $\lambda$ is the maximum work factor occurring in
our reduction lemmas, can be proven by induction on the size of
an instance.  The
work within each call is polynomial and does not add appreciably to the
overall time bound.

To determine the maximum work factor, we need to set a value for the
parameter $\epsilon$.  We used {\em Mathematica} to find a numerical value
of $\epsilon$ minimizing the maximum of the work factors involving
$\epsilon$, and found that for $\epsilon\approx 0.095543$
the work factor is $\approx 1.36443\approx\rec{4,4,5,5}$.  For $\epsilon$
near this value, the two largest work factors are
$\rec{3-\epsilon,4-\epsilon,4-\epsilon}$ (from
Lemma~\ref{lem:42-highdeg}) and
$\rec{1+\epsilon,4}$ (from Lemma~\ref{lem:32-highdeg}); the remaining work
factors are below 1.36.  The true optimum value of $\epsilon$ is thus the
one for which $\rec{3-\epsilon,4-\epsilon,4-\epsilon}=\rec{1+\epsilon,4}$.

As we now show, for this optimum $\epsilon$,
$\rec{3-\epsilon,4-\epsilon,4-\epsilon}=\rec{1+\epsilon,4}=\rec{4,4,5,5}$,
which also arises as a work factor in Lemma~\ref{lem:large3}.
Consider subdividing an instance of size $n$ into one of
size $n-(1+\epsilon)$ and another of size $n-4$, and then further
subdividing the first instance into subinstances of size
$n-(1+\epsilon)-(3-\epsilon)$, $n-(1+\epsilon)-(4-\epsilon)$, and
$n-(1+\epsilon)-(4-\epsilon)$. This four-way subdivision combines
subdivisions of type $\rec{1+\epsilon,4}$ and
$\rec{3-\epsilon,4-\epsilon,4-\epsilon}$,
so it must have a work factor between those two values.
But by assumption those two values equal each other, so they also
equal the work factor of the four-way subdivision,
which is just $\rec{4,4,5,5}$.
\end{proof}

We use the quantity $\rec{4,4,5,5}$ frequently in the remainder of the
paper, so we use $\Lambda$ to denote this value.
Theorem~\ref{thm:sss} immediately gives algorithms for some more well
known problems, some of which we improve later.
Of these, the least familiar is likely to be {\em list $k$-coloring}:
given at each vertex of a graph a list of $k$ colors chosen from some
larger set, find a coloring of the whole graph in which each vertex
color is chosen from the corresponding list~\cite{JenTof-95}.

\begin{corollary}
We can solve the 3-coloring and 3-list coloring problems in time
$O(\Lambda^n)$, the 3-edge-coloring
problem in time $O(\Lambda^m)$,
and the 3-SAT problem in time $O(\Lambda^t)$,
\end{corollary}

\begin{corollary}
There is a randomized algorithm which finds the solution to any
solvable $(d,2)$-CSP instance (with $d>3$) in expected time
$O((0.4518d)^n)$.
\end{corollary}

\begin{proof}
Randomly choose a subset of four values for each variable and apply our
algorithm to the resulting $(4,2)$-CSP problem.  Repeat with a new
random choice until finding a solvable $(4,2)$-CSP instance. The random
restriction of a variable has probability $4/d$ of preserving
solvability so the expected number of trials is $(d/4)^n$.
Each trial takes time $O(\Lambda^{(2-\epsilon)n})\approx O(1.8072^n)$.
The total expected time is therefore $O((d/4)^n 1.8072^n)$.
\end{proof}

\section{Vertex Coloring}

Simply by translating a 3-coloring problem into a $(3,2)$-CSP instance, as
described above, we can test 3-colorability in time $O(\Lambda^n)$.
We now describe some methods to reduce this time bound even further.

The basic idea is as follows: we find a small set of vertices $S\subset
V(G)$ with a large set $N$ of neighbors, and choose one of the $3^{|S|}$
colorings for all vertices in $S$.  For each such coloring, we translate
the remaining problem to a $(3,2)$-CSP instance.  The vertices in $S$ are
already colored and need not be included in the $(3,2)$-CSP instance.  The
vertices in $N$ now have a colored neighbor, so for each such vertex at
most two possible colors remain; therefore we can eliminate them from
the $(3,2)$-CSP instance using Lemma~\ref{lem:sss-elim}.
The remaining instance has $k=|V(G)-S-N|$ vertices,
and can be solved in time $O(\Lambda^k)$ by Theorem~\ref{thm:sss}.
The total time is thus $O(3^{|S|} \Lambda^k)$.
By choosing $S$ appropriately we can make this quantity smaller than
$O(\Lambda^n)$.

We can assume without loss of generality that all vertices in $G$ have
degree three or more, since smaller degree vertices can be removed
without changing 3-colorability.

As a first cut at our algorithm, choose $X$ to be any set of vertices,
no two adjacent or sharing a neighbor, and maximal with this property.
Let $Y$ be the set of neighbors of $X$.
We define a rooted forest $F$ covering $G$ as follows:
let the roots of $F$ be the vertices in $X$,
let each vertex in $Y$ be connected to its unique neighbor in $X$,
and let each remaining vertex $v$ in $G$ be connected to some neighbor
of $v$ in $Y$.  (Such a neighbor must exist or $v$ could have been added
to $X$).  We let the set $S$ of vertices to be colored consist of all of
$X$, together with each vertex in $Y$ having three or more children
in~$F$.

\begin{figure}[t]
$$\includegraphics[height=1.5in]{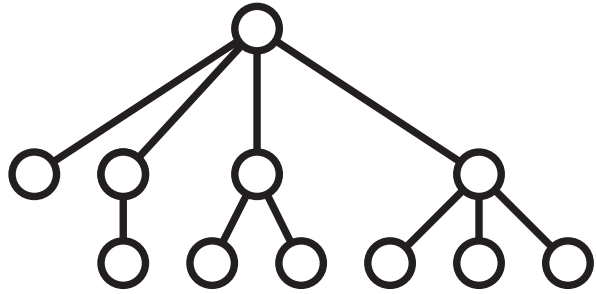}$$
\caption{Types of branch in a height-two tree (left-right): club,
stick, fork, broom.}
\label{fig:branches}
\end{figure}

\begin{figure}[t]
$$\includegraphics[height=1.5in]{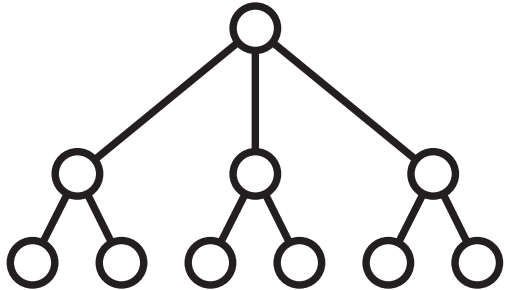}$$
\caption{Worst-case tree for 3-coloring algorithm: three forks.}
\label{fig:badtree}
\end{figure}

We classify the subtrees of $F$ rooted at vertices in $Y$ as follows
(Figure~\ref{fig:branches}).
If a vertex $v$ in $Y$ has no children, we call the subtree rooted at
$v$ a {\em club}.  If $v$ has one child, we call its subtree a {\em stick}.
If it has two children, we call its subtree a {\em fork}.
And if it has three or more children, we call its subtree a {\em broom}.

We can now compute the total time of our algorithm by multiplying
together a factor of $3$ for each vertex in $S$
(that is, the roots of the trees of $F$ and of broom subtrees) and a
factor of $\Lambda$ for each leaf in a stick or fork.  We define the
{\em cost} of a vertex in a tree $T$
to be the product $p$ of such factors involving vertices of $T$,
spread evenly among the vertices---if $T$ contains $k$ vertices the
cost is $p^{1/k}$.  The total time of the algorithm will then be $O(c^n)$
where $c$ is the maximum cost of any vertex.  It is not hard to show
that this maximum is achieved in trees consisting of three forks
(Figure~\ref{fig:badtree}), for
which the cost is $(3(\Lambda)^6)^{1/10}\approx 1.34488$.  Therefore we
can three-color any graph in time $O(1.34488^n)$.

We can improve this somewhat with some more work.

\subsection{Cycles of Degree-Three Vertices}

\begin{figure}[t]
$$\includegraphics[width=4in]{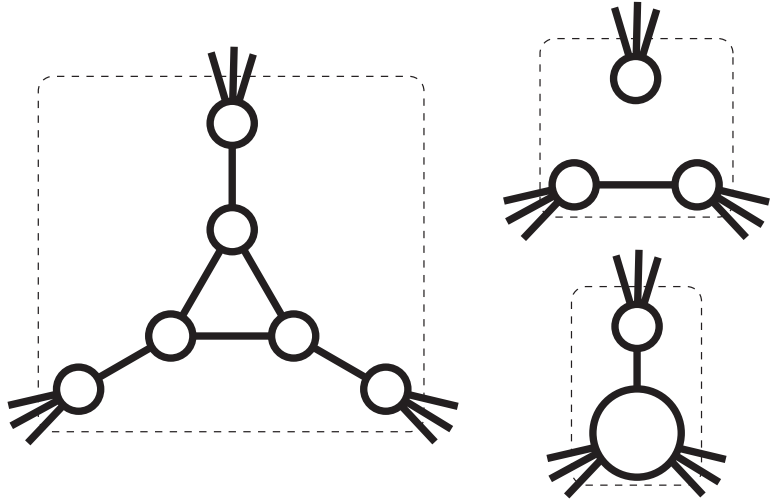}$$
\caption{Cases for elimination of a triangle of degree-three vertices
(left): add edge between two neighbors and eliminate triangle (top
right), or merge two neighbors and third triangle vertex into a single
supervertex (bottom right).}
\label{fig:triangle}
\end{figure}

\begin{figure}[t]
$$\includegraphics[width=5in]{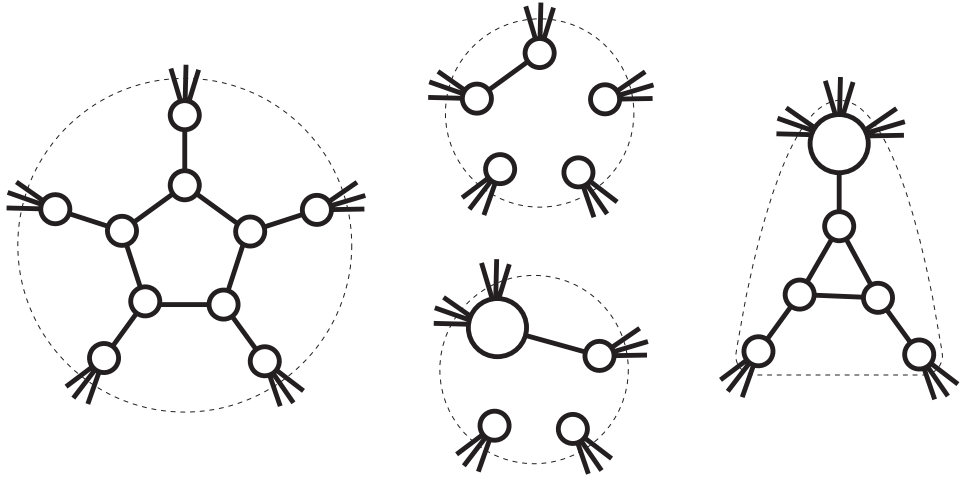}$$
\caption{Cases for elimination of a long odd cycle of degree-three
vertices (left): add edge between two neighbors and eliminate cycle
(top middle), merge two neighbors, add edge to third neighbor, and
eliminate cycle (bottom middle), or merge three neighbors and reduce
cycle length by two (right).}
\label{fig:pent}
\end{figure}

We begin by showing that we can assume that our graph has a special
structure: the degree-three vertices do not form any cycles.
For if they do form a cycle, we can remove it cheaply as follows.

\begin{lemma}\label{lem:3-cycle}
Let $G$ be a 3-coloring instance in which some cycle consists only of
degree-three vertices.  Then we can replace $G$ by smaller instances with
work factor at most $\rec{5,6,7,8}\approx 1.2433$.
\end{lemma}

\begin{proof}
Let the cycle $C$ consist of vertices $v_1$, $v_2$, $\ldots$, $v_k$.
We can assume without loss of generality that it has no chords,
since otherwise we could find a shorter cycle in $G$; therefore each
$v_i$ has a unique neighbor $w_i$ outside the cycle, although the $w_i$
need not be distinct from each other.

Note that, if any $w_i$ and $w_{i+1}$ are adjacent, then $G$ is
3-colorable iff $G\setminus C$ is; for, if we have a coloring of
$G\setminus C$, then we can color $C$ by giving $v_{i+1}$ the same color
as $w_i$, and then proceeding to color the remaining cycle vertices in
order $v_{i+2}$, $v_{i+3}$, $\ldots$,
$v_k$, $v_1$, $v_2$, $\ldots$, $v_i$.  Each successive vertex has only
two previously-colored neighbors, so there remains at least one free color
to use, until we return to $v_i$.  When we color $v_i$, all three of its
neighbors are colored, but two of them have the same color, so again
there is a free color.

As a consequence, if $C$ has even length, then $G$ is 3-colorable iff
$G\setminus C$ is; for if some $w_i$ and $w_{i+1}$ are given different
colors, then the above argument colors $C$, while if all $w_i$ have the
same color, then the other two colors can be used in alternation around
$C$.

The first remaining case is that $k=3$ (Figure~\ref{fig:triangle},
left).  Then we divide the problem into two smaller instances, by forcing
$w_1$ and
$w_2$ to have different colors in one instance (by adding an edge between
them, Figure~\ref{fig:triangle} top right) while forcing them to have the
same color in the other instance (by collapsing the two vertices into a
single supervertex,
Figure~\ref{fig:triangle} bottom right).  If we add an edge between
$w_1$ and $w_2$, we may remove
$C$, reducing the problem size by three. If we give them the same color
as each other, the instance is only colorable if $v_3$ is also given the
same color, so we can collapse $v_3$ into the
supervertex and remove the other two cycle vertices, reducing the problem
size by four. Thus the work factor in this case is
$\rec{3,4}\approx 1.2207$.

If $k$ is odd and larger than three, we form three smaller instances,
as shown in Figure~\ref{fig:pent}.
In the first, we add an edge between $w_1$ and $w_2$, and remove $C$,
reducing the problem size by $k$.  In the second, we collapse $w_1$ and
$w_2$, add an edge between the new supervertex and $w_3$, and again
remove $C$, reducing the problem size by $k+1$.  In the third instance,
we collapse $w_1$, $w_2$, and $w_3$.  This forces $v_1$ and $v_3$ to
have the same color as each other, so we also collapse those two
vertices into another supervertex and remove $v_2$, reducing the problem
size by four.  For $k\ge 7$ this gives work factor at most
$\rec{4,7,8}\approx 1.1987$.  For $k=5$ the subproblem with $n-4$
vertices contains a triangle of degree-three vertices, and can be further
subdivided into two subproblems of $n-7$ and $n-8$ vertices, giving the
claimed work factor.
\end{proof}

Any degree-three vertices remaining after the application of this lemma
must form components that are trees.  As we now show, we can also limit
the size of these trees.

\begin{lemma}\label{lem:bigtree}
Let $G$ be a 3-coloring instance containing a connected subset of eight
or more
degree-three vertices.  Then we can replace $G$ by smaller instances with
work factor at most $\rec{2,5,6}\approx 1.3247$.
\end{lemma}

\begin{proof}
Suppose the subset forms a $k$-vertex tree, and let $v$ be a vertex in
this tree such that each subtree formed by removing $v$ has at most
$k/2$ vertices.  Then, if $G$ is 3-colored, some two of the three
neighbors of $v$ must be given the same color, so we can split the
instance into three smaller instances, each of which collapses two of the
three neighbors into a single supervertex.  This collapse reduces the
number of vertices by one, and allows the removal of $v$ (since after
the collapse $v$ has degree two) and the subtree connected to the third
vertex.  Thus we achieve work factor $\rec{a,b,c}$ where
$a+b+c=k+3$ and $\max\{a,b,c\}\le k/2$.  The worst case is
$\rec{2,5,6}$, achieved when $k=8$ and the tree is a path.
\end{proof}

\subsection{Planting Good Trees}

We define a {\em bushy forest} to be an unrooted forest within a given
instance graph, such that each internal node has degree four or more
(for an example, see the top three levels of Figure~\ref{fig:pqrst}).
A bushy forest is {\em maximal} if
no internal node is adjacent to a vertex outside the forest,
no leaf has three or more neighbors
outside the forest, and no vertex outside the forest has four or more
neighbors outside the forest. If a leaf
$v$ does have three or more neighbors outside the forest, we could add
those neighbors to the tree containing
$v$, producing a bushy forest with more vertices.
Similarly, if a vertex outside the forest has four or more
neighbors outside the forest, we could extend the forest by adding
another tree consisting of that vertex and its neighbors.

As we now show, a maximal bushy forest must cover at least a constant
fraction of a 3-coloring instance graph.

\begin{lemma}\label{lem:const-frac-bushy}
Let $G$ be a graph in which all vertex degrees are three or more,
and in which there is no cycle of degree-three vertices, let $F$ be a
maximal bushy forest in $G$, and let $r$ denote the number of leaves in
$F$.  Then $|G\setminus F|\le 20r/3$.
\end{lemma}

\begin{proof}
Divide $G\setminus F$ into two subsets $X$ and $Y$,
where $X$ consists of the vertices of degree four or more
and $Y$ consists of the degree-three vertices.

Let $m_{A,B}$ denote the number of edges connecting sets $A$ and $B$.
Then each vertex in $X$ must have at least one edge connecting it to
$F$, and at most three edges connecting it to $Y$,
so $m_{X,F}\ge |X|$ and $m_{X,Y}\le 3|X|$.
Further, to avoid
cycles, each connected component in $Y$ must form a tree, and if such a
component has $k$ vertices, it must have $k+2$ edges leaving it,
and $k\le 7$ else we could apply Lemma~\ref{lem:bigtree}.
So, $m_{Y,X\cup F}\ge 9|Y|/7$.
If $3|X|\le 9|Y|/7$,
$m_{F,X\cup Y}=m_{F,X}+m_{F,X\cup Y}-m_{X,Y}\ge
9|Y|/7-2|X|=
3|X\cup Y|/10 + (69|Y|/70 - 23|X|/10)\ge 3|X\cup Y|/10$.
And if $3|X|\ge 9|Y|/7$, then again $m_{F,X\cup Y}\ge m_{F,X}\ge
|X|\ge 3|X\cup Y|/10$.

However, each leaf in $F$ has at most two edges outside $F$,
or $F$ would not be maximal, so $|X\cup Y|\le 10m_{F,X\cup Y}/3\le
20r/3$.
\end{proof}

\subsection{Pruning Bad Trees}

\begin{figure}[p]
$$\includegraphics[width=3.25in]{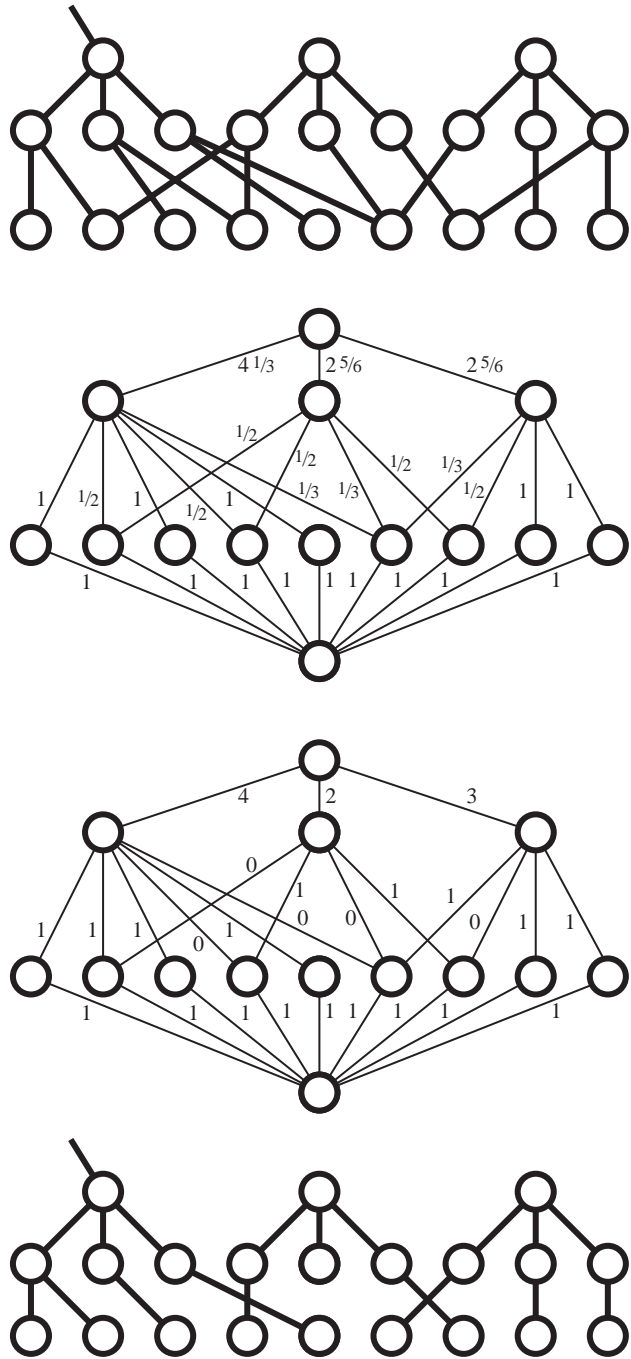}$$
\caption{Use of maximum flow to find a good height-two forest. Top:
forest $T$ of $K_{1,3}$ subgraphs and adjacent vertices in $Y$. Top
middle: flow graph and fractional flow formed by dividing flow equally at
each vertex in $Y$. The edge capacities are all one, except for the top
three which are respectively 5, 3, and 3.  Bottom middle: maximum integer
flow for the same flow graph.  Bottom: height-two forest corresponding to
the given integer flow.}
\label{fig:flow}
\end{figure}

After finding a maximal bushy forest $F$,
we find a second forest $H$ in the remaining graph $G\setminus F$,
as follows.  Note that, due to the maximality of $F$,
each vertex in $G\setminus F$ has at most three neighbors in $G\setminus
F$. We first choose a maximal set $T$ of disjoint $K_{1,3}$
subgraphs in $G\setminus F$.
Then, we increase the size of $T$ as much as possible by operations in
which we remove one $K_{1,3}$ from $T$ and form two
$K_{1,3}$ subgraphs from the remaining vertices.

Let $X$ denote the set of vertices in $G\setminus(T\cup F)$
that are adjacent to vertices in $F$.  By the maximality of $F$,
each vertex in $F$ is adjacent to at most two vertices in $X$.
Let $Y=G\setminus(X\cup T\cup F)$ denote the remaining vertices.
By the maximality of $T$, each vertex in $Y$ is adjacent to at most two
vertices in $X\cup Y$, and so must have a neighbor in $T$.
Since $G\setminus F$ contains no degree-four vertices,
each vertex in $T$ must have at most two neighbors in $Y$.
As we now show, we can assign vertices in $Y$ to trees in $T$,
extending each tree in $T$ to a tree of height at most two,
in such a way that we do not form any tree with three forks,
which would otherwise be the worst case for our algorithm.

\begin{lemma}
Let $F$, $T$, $X$, and $Y$ be as above.  Then there exists a forest $H$
of height two trees with three branches each, such that the vertices
of $H$ are exactly those of $S\cup Y$, such that each tree in $H$
has at most five grandchildren, and such that any tree with four or
more grandchildren contains at least one vertex with degree four or more
in $G$.
\end{lemma}

\begin{proof}
We first show how to form a set $H'$ of non-disjoint trees in $T\cup Y$,
and a set of weights on the grandchildren of these trees, such that
each tree's grandchildren have weight at most five.

To do this, let each tree in $H'$ be formed by one of the $K_{1,3}$
trees in $T$, together with all possible grandchildren in $Y$ that are
adjacent to the $K_{1,3}$ leaves.  We assign each vertex in $Y$ unit
weight, which we divide equally among the trees it belongs to.

Then, suppose for a contradiction that some tree $h$ in $H'$ has
grandchildren with total weight more than five.  Then, its grandchildren
must form three forks, and at least five of its six grandchildren must
have unit weight; i.e., they belong only to tree $h$.  Note that each
vertex in $Y$ must have degree three, or we could have added it to the
bushy forest, and all its neighbors must be in $S\cup Y$, or we could
have added it to
$X$. The
unit weight grandchildren each have one neighbor
in $h$ and two other neighbors in $Y$.  These two other neighbors must be
one each from the two other forks in $h$, for, if to the contrary some
unit-weight grandchild $v$ does not have neighbors in both forks, we could
have increased the number of trees in $T$ by removing
$h$ and adding new trees rooted at $v$ and at the missed fork.

Thus, these five grandchildren each connect to two other grandchildren,
and (since no grandchild connects to three grandchildren) the six
grandchildren together form a degree-two graph, that is, a union of
cycles of degree-three vertices. But after applying
Lemma~\ref{lem:3-cycle} to
$G$, it contains no such cycles.  This contradiction implies that the
weight of $h$ must be at most five.

Similarly, if the weight of $h$ is more than three, it must have at
least one fork, at least one unit-weight grandchild outside that fork,
and at least one edge connecting that grandchild to a grandchild within
the fork.  This edge together with a path in $h$ forms a cycle, which
must contain a high degree vertex.

We are not quite done, because the assignment of grandchildren to trees
in $H'$ is fractional and non-disjoint.  To form the desired forest $H$,
construct a network flow problem in which the flow source is connected
to a node representing each tree $t\in T$ by an edge with capacity
$w(t)=5$ if $t$ contains a high degree vertex and
capacity $w(t)=3$ otherwise.  The node corresponding to tree $t$ is
connected by unit-capacity edges to nodes corresponding to the vertices in
$Y$ that are adjacent to $t$, and each of these nodes is connected by
a unit-capacity edge to a flow sink.
Then the fractional weight system above defines a flow that saturates all
edges into the flow sink and is therefore maximum 
(Figure~\ref{fig:flow}, middle top). But any maximum flow problem with
integer edge capacities has an integer solution (Figure~\ref{fig:flow},
middle bottom).  This solution must continue to saturate the sink edges,
so each vertex in $Y$ will have one unit of flow to some tree $t$,
and no flow to the other adjacent trees.
Thus, the flow corresponds to an assignment of vertices in $Y$ to adjacent
trees in
$T$ such that each tree is assigned at most
$w(t)$ vertices.  We then simply let each tree in $H$ consist of a tree
in $T$ together with its assigned vertices in $Y$ (Figure~\ref{fig:flow},
bottom).
\end{proof}

\subsection{Improved Tree Coloring}

\begin{figure}[t]
$$\includegraphics[width=5in]{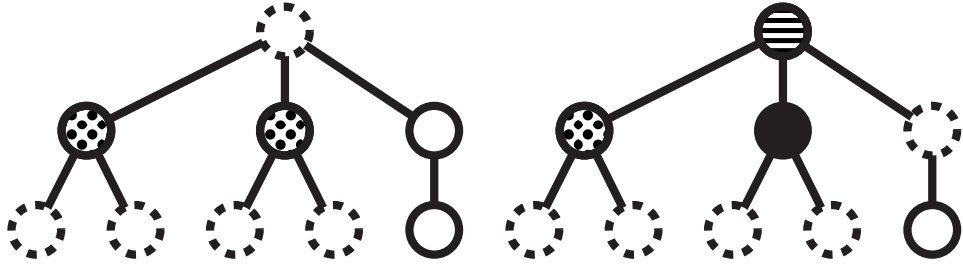}$$
\caption{Coloring a tree with two forks and one stick. If the two fork
vertices are colored the same (left), five neighbors (dashed) are
restricted to two colors, leaving the two stick vertices for the
$(3,2)$-CSP instance. If the two forks are colored differently (right),
they force the tree root to have the third color, leaving only one vertex
for the $(3,2)$-CSP instance.}
\label{fig:twoforks}
\end{figure}

We now discuss how to color the trees in the height-two forest $H$
constructed in the previous subsection. As in the discussion at the start
of this section, we color some vertices (typically just the root) of each
tree in
$H$, leave some vertices (typically the grandchildren) to be part of a
later
$(3,2)$-CSP instance, and average the costs over all the vertices in the
tree.  However, we average the costs in the following strange way:
a cost of $\Lambda$ is assigned to any vertex with degree four
or higher in $G$, as if it was handled as part of the $(3,2)$-CSP
instance.  The remaining costs are then divided equally among the
remaining vertices.

\begin{lemma}\label{lem:two-fork}
Let $T$ be a tree with three children and at most five grandchildren. 
Then
$T$ can be colored with cost per degree-three vertex at most
$(3\Lambda^3)^{1/7}\approx 1.3366$.
\end{lemma}

\begin{proof}
First, suppose that $T$ has exactly five grandchildren.
At least one vertex of $T$ has high degree.
Two of the children $x$ and $y$ must be the roots of forks,
while the third child $z$ is the root of a stick.
We test each of the nine possible colorings of $x$ and $y$.
In six of the cases, $x$ and $y$ are different, forcing the root
to have one particular color (Figure~\ref{fig:twoforks}, right).
In these cases the only remaining vertex after translation to a
$(3,2)$-CSP instance and application of Lemma~\ref{lem:sss-elim} will be
the child of $z$, so in each such case $T$ accumulates a further cost of
$\Lambda$.  In the three cases in which $x$ and $y$ are colored the
same (Figure~\ref{fig:twoforks}, left), we must also take an additional
factor of
$\Lambda$ for
$z$ itself.  One of these $\Lambda$ factors goes to a high degree
vertex, while the remaining work is split among the remaining eight
vertices. The cost per vertex in this case is then at most
$(6 + 3\Lambda)^{1/8}\approx 1.3351$.

If $T$ has fewer than five grandchildren, we choose a color for the root
of the tree as described at the start of the section.
The worst case occurs when the number of grandchildren is either
three or four, and is
$(3\Lambda^3)^{1/7}\approx 1.3366$.
\end{proof}

\subsection{The Vertex Coloring Algorithm}

\begin{figure}[t]
$$\includegraphics[width=5in]{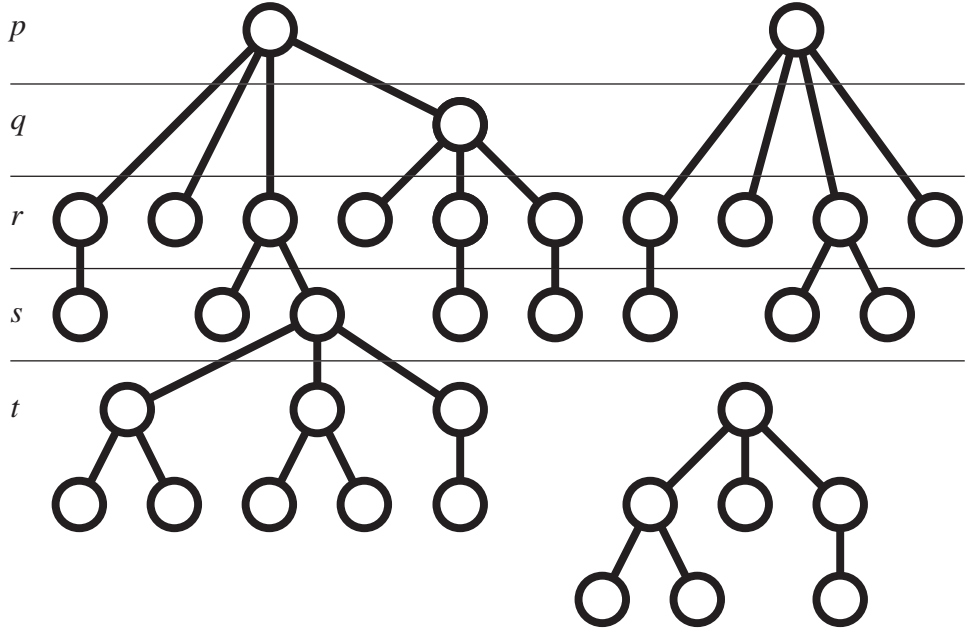}$$
\caption{Partition of vertices into five groups: $p$ bushy forest
roots, $q$ other bushy forest internal nodes, $r$ bushy forest leaves,
$s$ vertices adjacent to bushy forest leaves, and $t$ degree-three
vertices in height-two forest.}
\label{fig:pqrst}
\end{figure}

\begin{theorem}
We can solve the 3-coloring problem in time
$O((2^{3/49} 3^{4/49} \Lambda^{24/49})^n)
\approx 1.3289^n$.
\end{theorem}

\begin{proof}
As described in the preceding sections, we find a maximal bushy forest,
then cover the remaining vertices by height-two trees.
We choose colors for each internal vertex in the bushy forest,
and for certain vertices in the height-two trees as described
in Lemma~\ref{lem:two-fork}.  Vertices adjacent to these colored vertices
are restricted to two colors, while the remaining vertices form
a $(3,2)$-CSP instance and can be colored using our general
$(3,2)$-CSP algorithm.
Let $p$ denote the number of vertices that are roots in the bushy
forest; $q$ denote the number of non-root internal vertices;
$r$ denote the number of bushy forest leaves; $s$ denote the number
of vertices adjacent to bushy forest leaves; and $t$ denote the number
of remaining vertices, which must all be degree-three vertices in the
height-two forest (Figure~\ref{fig:pqrst}). Then the total time for the
algorithm is at most
$3^p 2^q \Lambda^s (3\Lambda^3)^{t/7}$.

We now consider which values of these parameters give the worst
case for this time bound, subject to the constraints
$p,q,r,s,t\ge 0$, $p+q+r+s+t=n$,
$4p+2q\le r$ (from the definition of a bushy forest),
$2r\ge s$ (from the maximality of the forest), and
$20r/3\ge s+t$ (Lemma~\ref{lem:const-frac-bushy}).
We ignore the slightly tighter constraint $p\ge 1$ since
it only complicates the overall solution.

Since the work per vertex in $s$ and $t$ is larger than that in the bushy
forests, the time bound is maximized when $s$ and $t$ are as large as
possible; that is, when $s+t=20r/3$. Further since the work per vertex
in $s$ is larger than that in $t$, $s$ should be as large as possible;
that is, $s=2r$ and $t=14r/3$. Increasing
$p$ or
$q$ and correspondingly decreasing $r$, $s$, and $t$ only increases the
time bound, since we pay a factor of 2 or more per vertex in $p$ and $q$
and at most $\Lambda$ for the remaining vertices, so in the worst
case the constraint $4p+2q\le r$ becomes an equality.

It remains only
to set the balance between parameters $p$ and $q$.  There are
two candidate solutions: one in which $q=0$,
so $r=4p$, and one in which $p=0$, so $r=2q$.
In the former case $n=p+4p+8p+56p/3=95p/3$
and the time bound
is $3^p \Lambda^{8p} (3\Lambda^3)^{8p/3}
=3^{11p/3}\Lambda^{16p}\approx 1.3287^n$.
In the latter case
$n=q+2q+4q+28q/3=49q/3$
and the time bound is
$2^q \Lambda^{4q} (3\Lambda^3)^{4q/3}
=2^q 3^{4q/3} \Lambda^{8q}
\approx 1.3289^n$.
\end{proof}

\section{Edge Coloring}

We now describe an algorithm for finding edge colorings of undirected
graphs, using at most three colors, if such colorings exist.
We can assume without loss of generality that the graph has vertex
degree at most three.
Then $m\le 3n/2$, so by applying our vertex coloring algorithm to the
line graph of $G$ we could achieve time bound
$1.3289^{3n/2}\approx 1.5319^n$.  Just as we improved our vertex coloring
algorithm by performing some reductions in the vertex coloring model
before treating the problem as a $(3,2)$-CSP instance, we improve this
edge coloring bound by performing some reductions in the edge coloring
model before treating the problem as a vertex coloring instance.

\begin{figure}[t]
$$\includegraphics[width=3.5in]{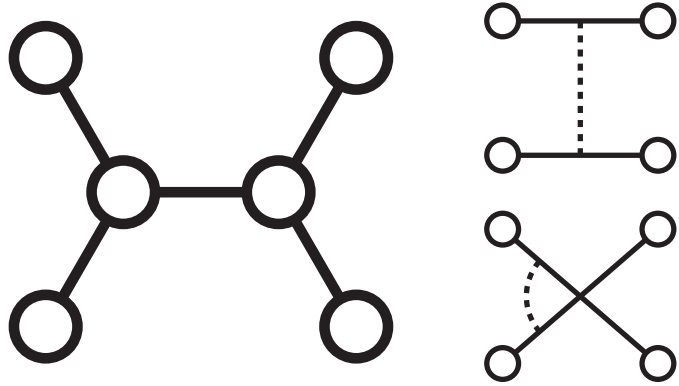}$$
\caption{Replacement of five edges (left) by two constrained edges
(right).}
\label{fig:splice}
\end{figure}

The main idea is to solve a problem intermediate in generality
between 3-edge-coloring and 3-vertex-coloring: 3-edge-coloring with some
added constraints that certain pairs of edges should not be the same
color.

\begin{lemma}\label{lem:splice}
Suppose a constrained 3-edge-coloring instance contains an unconstrained
edge connecting two degree-three vertices.
Then the instance can be replaced by two smaller instances with three
fewer edges and two fewer vertices each.
\end{lemma}

\begin{proof}
Let the given edge be $(w,x)$,
and let its four neighbors be $(u,w)$, $(v,w)$, $(x,y)$, and $(x,z)$.
Then $(w,x)$ can be colored only if its four neighbors
together use two of the three colors, which forces these neighbors to be
matched into equally colored pairs in one of two ways.  Thus, we can
replace the instance by two smaller instances: one
in which we replace the five edges by the two edges $(u,y)$ and $(v,z)$,
and one in which we replace the five edges by the two edges $(u,z)$ and
$(v,y)$; in each case we add a constraint between the two new edges.
\end{proof}

The reduction operation described in Lemma~\ref{lem:splice}
is depicted in Figure~\ref{fig:splice}. 

We let $m_3$ denote the number of edges with three neighbors in an
unconstrained 3-edge-coloring instance, and $m_4$ denote the number of
edges with four neighbors.  Edges with fewer neighbors can be removed at
no cost, so we can assume without loss of generality that $m=m_3+m_4$.

\begin{lemma}\label{lem:many-splices}
In an unconstrained 3-edge-coloring instance, we can find in polynomial
time a set $S$ of
$m_4/3$ edges such that Lemma~\ref{lem:splice} can be applied
independently to each edge in $S$.
\end{lemma}

\begin{proof}
Use a maximum matching algorithm in the graph induced by the edges
with four neighbors.  If the graph is 3-colorable, the resulting
matching must contain at least $m_4/3$ edges.  Applying
Lemma~\ref{lem:splice} to an edge in a matching neither constrains any
other edge in the matching, nor causes the remaining edges to stop being
a matching.
\end{proof}

\begin{lemma}\label{lem:m3}
$m_3 = \frac{6}{5}n - \frac{4}{5}m_4$.
\end{lemma}

\begin{proof}
Assign a charge of $6/5$ to each vertex of the graph,
and redistribute this charge equally to each incident edge.
Further assign an additional $1/5$ charge to each four-neighbor
edge.  Then each edge receives a unit charge, so
$m_3+m_4=m=(6/5)n+(1/5)m_4$.
Subtracting $m_4$ from both sides yields the result.
\end{proof}

\begin{theorem}
We can 3-edge-color any 3-edge-colorable graph, in time
$O(2^{n/2})$.
\end{theorem}

\begin{proof}
We apply Lemma~\ref{lem:many-splices}, resulting in a 
set of $2^{m_4/3}$ constrained
3-edge-coloring problems each having only $m_3$ edges.  We then treat
these remaining problems as 3-vertex-coloring problems on the
corresponding line graphs, augmented by additional edges
representing the constraints added by Lemma~\ref{lem:splice}.
The time for this algorithm is thus at most
$O(1.3289^{m_3} 2^{m_4/3})$.
By Lemma~\ref{lem:m3},
we can rewrite this bound as
$O(1.3289^{6n/5}(2^{1/3}1.3289^{-4/5})^{m_4})$.
Since $2^{1/3}1.3289^{-4/5}>1$,
this time bound is maximized when $m_4$
is maximized, which occurs when $m_4=3n/2$ and $m_3=0$.
For this value, all the work occurs within Lemma~\ref{lem:many-splices},
and gives the stated time bound.
\end{proof}

\paragraph{Acknowledgments}
A preliminary version of this paper was presented at the 36th IEEE
Symp. Foundations of Comp. Sci., 1995. The first author thanks
Russell Impagliazzo and Richard Lipton for bringing this problem to his
attention.  Both authors thank Laszlo Lovasz for helpful discussions.

\bibliographystyle{abuser}
\bibliography{3color}

\end{document}